\documentclass[11pt,a4paper]{article}

\usepackage{a4wide}
\usepackage{amsmath}
\usepackage{amssymb}
\usepackage{booktabs}
\usepackage{graphicx}
\usepackage{hyperref}
\usepackage{xcolor}

\newcommand{\fref}[1]{Fig.~\ref{#1}}
\newcommand{\Fref}[1]{Figure~\ref{#1}}

\newcommand{\GeV}{\ensuremath{\,\mathrm{GeV}}}
\newcommand{\TeV}{\ensuremath{\,\mathrm{TeV}}}
\newcommand{\alphas}{\ensuremath{\alpha_\mathrm{S}}}

\begin{document}

\title{\textbf{Standard Model physics at the LHC}}
\author{Jan Kretzschmar\\University of Liverpool
}
\date{}
\maketitle

\begin{abstract}
In this contribution to the volume ``From My Vast Repertoire -- The
Legacy of Guido Altarelli'' I discuss selected topics of Standard Model (SM)
physics at the LHC and their connection to the work of Guido
Altarelli. An overarching connection is given by Guido's fundamental
contributions to our understanding of the strong force, parton distribution
functions and their
evolution, and techniques to calculate theoretical predictions through
perturbative Quantum Chromodynamics.
The increasingly precise LHC measurements of diverse
processes can be confronted with SM predictions relying on these
foundations to test the SM and contribute to the
knowledge of the proton structure.
The LHC experiments have also started to perform
precise measurements of SM parameters, such as the mass of
the $W$ boson, and have measured a variety of processes sensitive to
interactions of multiple electroweak bosons.
The discovery of the Higgs boson in 2012 and the
measurements of its properties are crucial to establish the
nature of the mechanism of electroweak symmetry breaking and the status
of these studies is summarised.
\end{abstract}
\vspace{0.5em}

\section{Introduction}
\label{sec:lhc:intro}

Before the year 2009, the start of the Large Hadron Collider (LHC) was
eagerly anticipated in the whole particle physics community. As Guido
Altarelli put it: ``physics [was] in a deadlock and [needed] to be
restarted.''~\cite{Altarelli:2011vt}. The Standard Model of particle
physics (SM) was by many seen just a low-energy effective version of the
ultimate theory. The LHC was expected to take our understanding to the
next level. In the following I attempt to discuss where the endeavour
to explore the SM at the energy frontier has lead us in the last seven
years.

The LHC~\cite{Evans:2008zzb} is the most
powerful accelerator ever built. From 2010 to 2012 it provided
proton--proton ($pp$) collisions at an unprecedented centre-of-mass
energy of $\sqrt{s} = 7$--$8\TeV$ in Run 1. In 2015 the LHC started
the Run 2 operation at $\sqrt{s} = 13\TeV$, close to the design value
of $14\TeV$. With this, it exceeds the typical energy scale of
collisions at the previous generation of colliders, LEP, HERA and
TeVatron, by about an order of magnitude. The LHC now regularly
delivers an instantaneous luminosity higher by 50\% than the original
design goal of $10^{34}\,\mathrm{cm}^{-1}\,\mathrm{s}^{-1}$ to the
ATLAS and CMS experiments~\cite{ATLAS, CMS}, exceeding previous
colliders by about two orders of magnitude.

The SM describes the interactions of three generations of quarks and
leptons through the gauge fields of the electroweak and strong
interactions, with the electroweak symmetry being spontaneously broken
by the Englert-Brout-Higgs mechanism. The unprecedented collision
energy and large data sets provided by the LHC allow to probe the
fundamental building blocks of matter and their interactions at a new
level. It was certainly one of the ``major physics discoveries of our
time''~\cite{Altarelli:2013lla} when the LHC experiments discovered
``a'' Higgs boson in 2012~\cite{Aad:2012tfa,Chatrchyan:2012xdj}.  The
Higgs boson was the final of the SM particles and was seen as
``directly related to most of the major open problems of particle
physics''~\cite{Altarelli:2011vt}.

It is important to remember, that the $pp$ collisions delivered by the
LHC are complex interactions of composite, strongly-interacting
particles. Without the framework provided by the SM, we would hardly
be in a situation to exploit the data at such a detailed level as we
do it these days: decades of efforts in theory and phenomenology
enable us to connect the basic Lagrangian to Feynman diagrams and those to
quantitative predictions of cross sections and decay rates.  A
striking feature of the LHC data is the vast spectrum of measurements
(and searches) that are possible, thereby testing a wide diversity of
processes as well as very different energy scales.

\section{The role of QCD in LHC physics}
\label{sec:lhc:qcd}

The protons collided by the LHC are strongly-interacting particles
composed of quarks and gluons. It is thus somewhat obvious, that the
theory of strong interactions, Quantum Chromodynamics (QCD), is of the
highest importance for LHC physics. This is also true, if the aim is
to study electroweak interactions.  Or, as Guido Altarelli put it:
``the understanding QCD processes is an essential prerequisite for all
possible discoveries'' at the LHC~\cite{Altarelli:2010uu}.
Thanks to the work of Guido Altarelli and many others at the end
of the 1970s, QCD was established as the theory of strong interactions
and the framework of perturbative calculations was worked out. At the
level of the Lagrangian ``QCD is a 'simple' theory, but actually this
simple theory has an extremely rich dynamical
content''~\cite{Altarelli:2010uu}.

The calculation of a cross section can be factorized into a part
describing the structure of hadrons in terms of the Parton
Distribution Functions (PDFs) and another part describing the
interactions of the fundamental quarks and gluons. The PDFs are not
static objects, but evolve with the energy scale of the interaction, a
property studied by Dokshitzer, Gribov and
Lipatov~\cite{Gribov:1972ri, Gribov:1972rt, Dokshitzer:1977sg}, and
then rederived by Altarelli and Parisi ``in a direct way from
the basic vertices of QCD''~\cite{Altarelli:1977zs}. The resulting
evolution equations are known as DGLAP equations and form
a cornerstone of our understanding of QCD. The formulation given by
Guido Altarelli has allowed to systematically improve the
precision. The splitting functions that describe the evolution are
nowadays known to the three-loop level~\cite{Vogt:2004mw,Moch:2004pa}
and partially beyond this.

The elementary cross sections can be systematically calculated and
similarly improved by an expansion in the strong coupling constant
\alphas. Guido Altarelli contributed to one of the first higher order
calculations~\cite{Altarelli:1979ub}: the production of di-leptons in
hadron collisions through the Drell--Yan
process~\cite{Drell:1970wh}. The Drell--Yan process will be further
discussed in Section~\ref{sec:lhc:dy}. The mechanism to improve
the calculations by considering diagrams higher-order in \alphas\ has
been critical. For LHC physics purposes it
has turned out, that the next-to-leading order (NLO) is the minimum to
have a reliable prediction, but often the
next-to-next-to-leading order (NNLO) still provides an important
correction.

Guido Altarelli's contributions have such paved the way to
quantitative predictions for QCD processes. The exploitation of
the large LHC data sets keeps increasing the demands of
experimentalists on the theory accuracy, which typically reaches the
$5$--$10\%$ percent level and may reach below $1\%$ for simple
processes. Still, the precision in the measurements is often better
than the theory uncertainties, limiting the interpretation of the
data.

\section{A broad overview of SM processes studied at the LHC}
\label{sec:lhc:overview}

\begin{figure}
\centerline{\includegraphics[width=0.98\linewidth]{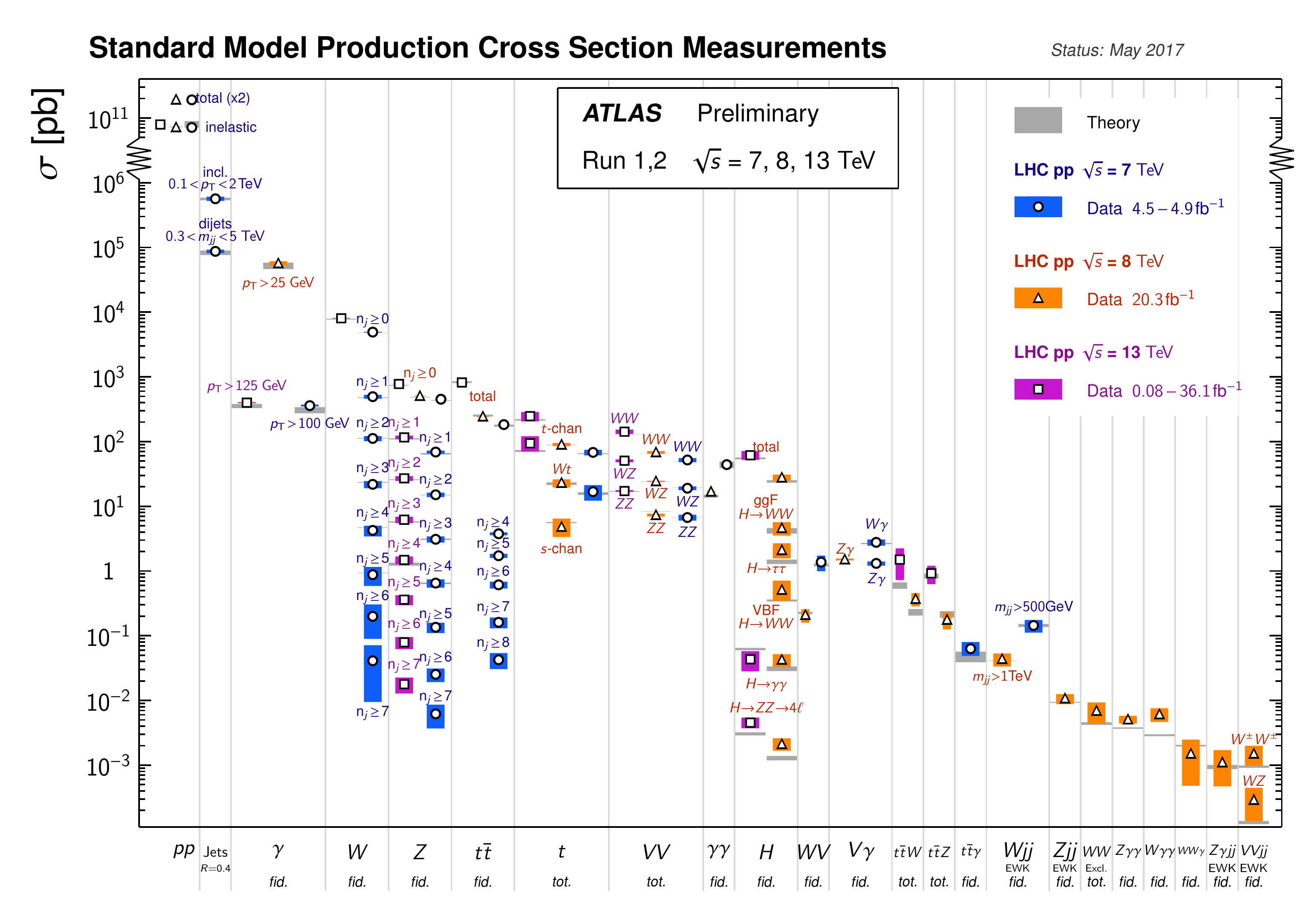}}
\caption{A summary of cross-section measurements in $pp$ collisions at
  $\sqrt{s}=7,8,13\TeV$ for a variety of SM processes by the ATLAS
  collaboration~\cite{summaryATLAS}. The measurements are compared to
  the respective SM theory predictions. A similar summary is available
  for results by the CMS collaboration~\cite{summaryCMS}.}
\label{fig:lhc:overview}
\end{figure}

The diversity of processes and energy scales accessible at the LHC is
exceptionally wide. A very much simplified summary of cross-section
measurements is given in \fref{fig:lhc:overview}, where each measured
point often stands for a complex, multi-differential analysis and
typically $10$--$1000$ differential measurements. The range of
cross sections studied ranges from $100\,\mathrm{mb}$ for the total
$pp$ cross section to $1\,\mathrm{fb}$ for the rarest processes, thus
spanning $14$ orders of magnitude.~\footnote{In addition, this
  overview does not even attempt to capture many other important SM
  measurements performed with LHC data, as for example the study of
  rare or CP-violating $B$-hadron decays with the
  LHCb~\cite{Alves:2008zz}, CMS and ATLAS experiments or the study of
  Heavy Ion collisions and the quark--gluon plasma, where the major
  contributions of the ALICE experiments lie~\cite{Aamodt:2008zz}.}

\begin{figure}
  \includegraphics[width=0.5\linewidth]{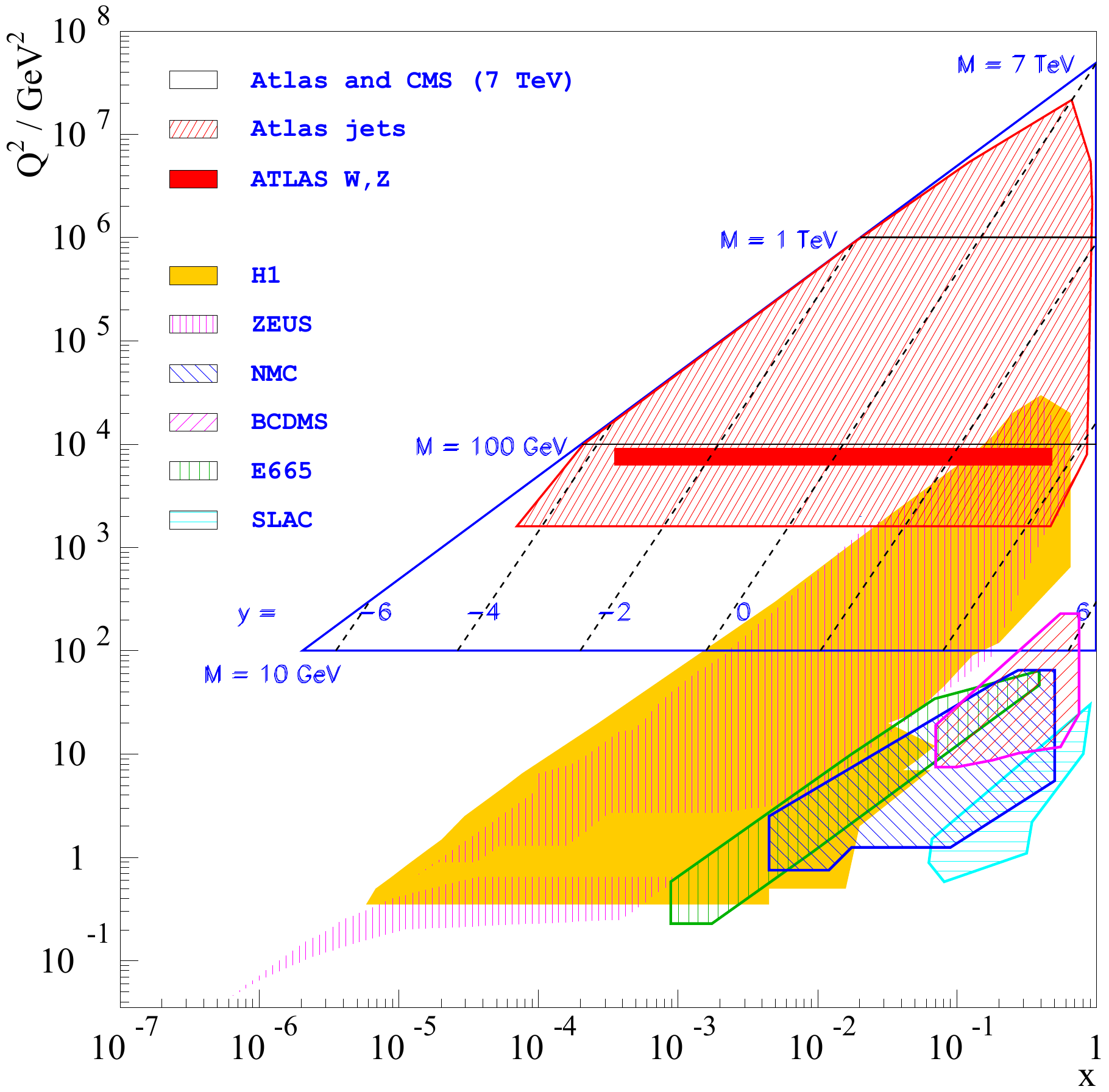}%
  \hspace{.2cm}%
  \includegraphics[width=0.45\linewidth]{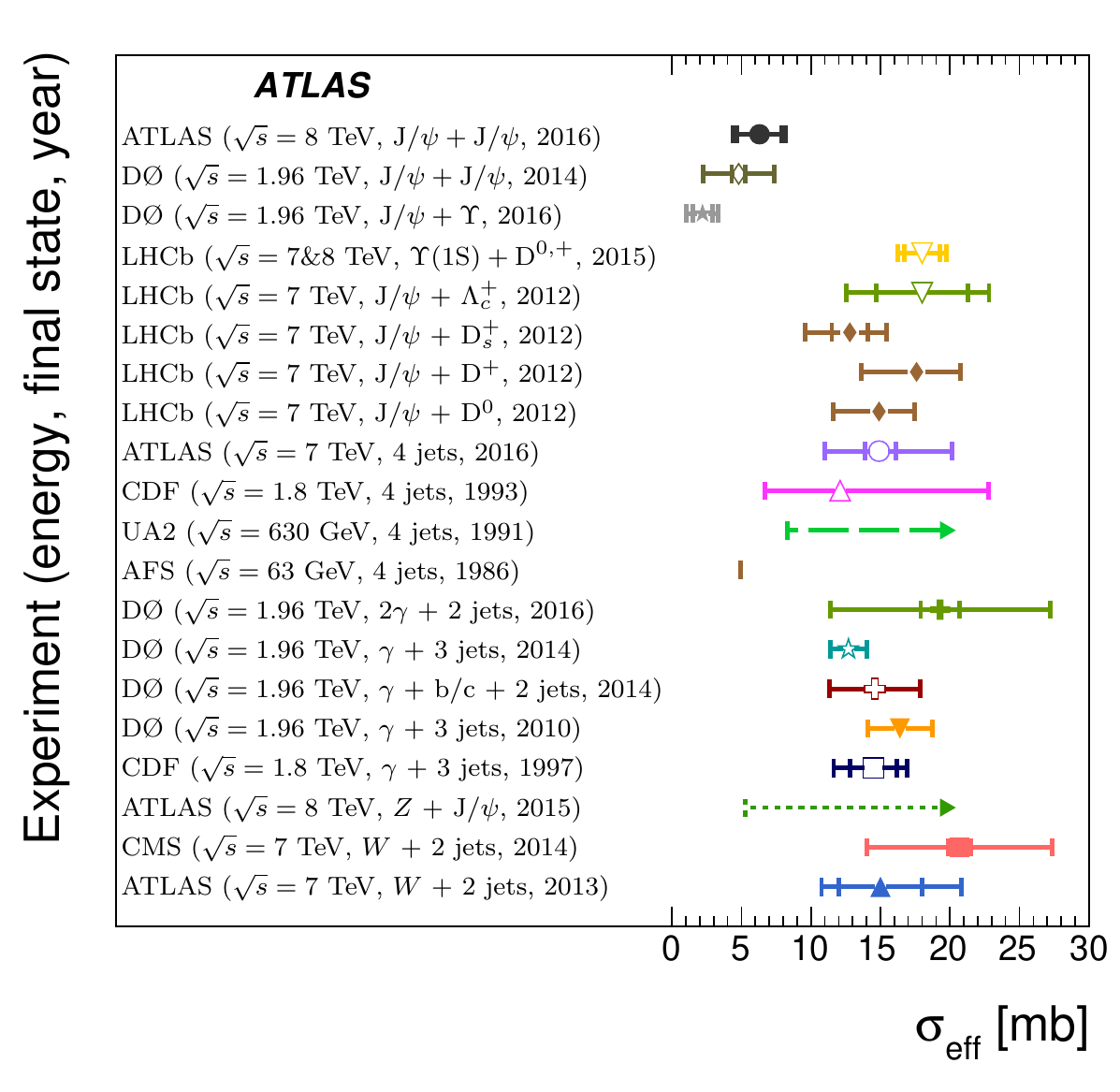}
  \caption{(left) Illustration of the plane in Bj\"{o}rken-$x$ and scale
    $Q^2$ relevant for the determination of PDFs from DIS data by the
    HERA experiments (H1, ZEUS) and a selection of fixed-target
    experiments (NMC, BCDMS, E665, SLAC). The kinematic region
    accessed by the LHC in terms of mass $M$ and rapidity $y$ of the
    produced system is shown for $\sqrt{s}=7\TeV$ together with
    example coverage by specific processes (jet and W,Z
    production). (right) A summary of effective DPI cross sections
    $\sigma_\mathrm{eff}$ derived from a variety of processes studied at
    different hadron-collider experiments~\cite{Aaboud:2016fzt}.}
  \label{fig:lhc:xq2}
\end{figure}

The theoretical predictions for the vast majority of the processes
displayed in \fref{fig:lhc:overview} use the techniques of
perturbative QCD. The necessary PDF information is largely derived
from deep-inelastic lepton--nucleon scattering (DIS), although more
recently the information from LHC measurements have started to become
useful to constrain PDFs further. \fref{fig:lhc:xq2}(left) illustrates
the relevant range of the Bj\"{o}rken-$x$ momentum fraction for LHC
physics, which is roughly from $10^{-4}$ to $1$. Indeed DIS
fixed-target experiments and the HERA $ep$ collider cover this range,
however a significant evolution in energy scale is required to reach
the LHC region. The broad agreement of the measurements with the
predictions is thus a major triumph for the SM in general and QCD and the
DGLAP formalism more specifically.

In the following, I will discuss a selection of recent measurements,
starting with the most abundant processes involving only the
strong interaction, moving on to processes involving the production of
single electroweak bosons and a small selection of multi-boson
processes before concluding with the current status of the studies of
the Higgs boson.

\section{Soft QCD Processes}
\label{sec:lhc:softqcd}

The majority of $pp$ interactions take place at small energy scales,
where \alphas\ is large and non-perturbative effects dominate.
Effects of this ``soft regime'', such as inelastic minimum-bias
interactions, often cannot be systematically computed through the
usual perturbative QCD techniques. We thus need to resort to
phenomenological models integrated in Monte Carlo (MC) tools like
Pythia, Herwig or Epos~\cite{Sjostrand:2007gs, Bellm:2015jjp,
  Pierog:2013ria} that are incorporate aspects of QCD and require
tuning to data. Despite the fact that these MC tools are not directly
connected to the SM Lagrangian, they are indispensable to the work of
experimentalists. MC simulations are used to model the
soft effects and their influence on measurements of processes at large
energy scale, such as hadronisation of quarks and gluons and
additional interactions beyond the hard scattering (``underlying
event''). It remains a challenge to develop and constrain these models
further such that non-perturbative effects will not limit the
interpretation of LHC measurements in the future. Detailed
measurements often show significant discrepancies to the model
predictions. For example, the ALICE collaboration recently reported an
enhanced production of strange hadrons in high-multiplicity $pp$
collisions~\cite{ALICE:2017jyt}, suggesting a lack of
understanding of the mechanism in the current MC models.

Another interesting aspect is the existence of hard double-parton
interactions (DPI), which despite taking place at perturbative
energy scales point beyond
our typical pQCD framework. In a simple model, the DPI interaction may
factorize into two effectively independent parton--parton scatters
within the same $pp$ interaction, however the probability for the second
interaction is adjusted by a factor
$\sigma_\mathrm{inel}/\sigma_\mathrm{eff}$. Here
$\sigma_\mathrm{inel}$ refers to the total inelastic interaction cross
section and $\sigma_\mathrm{eff}$ is the so-called effective DPI cross
section, a parameter to be determined in the experiment. In
\fref{fig:lhc:xq2}(right) an overview of measurements of
$\sigma_\mathrm{eff}$ from both TeVatron and LHC analyses is given.
The current data set is still sparse and does not allow to
determine a clear dependence on the centre-of-mass energy or the final
states considered. Most analyses find a value for $\sigma_\mathrm{eff}$ that is a
significant factor $\sim 5$ below $\sigma_\mathrm{inel}$. This possibly
suggests a ``hot-spot'' interpretation, where the existence of a first hard
interaction increases the probability to find a second hard
scatter. Interestingly the values of $\sigma_\mathrm{eff}$
extracted from double quarkonia final states feature further reduced
$\sigma_\mathrm{eff}$ values, suggesting an enhanced correlation in
these cases.

\section{Production of jets and heavy-flavour jets}
\label{sec:lhc:jet}

\begin{figure}[th]
  \includegraphics[width=0.48\linewidth]{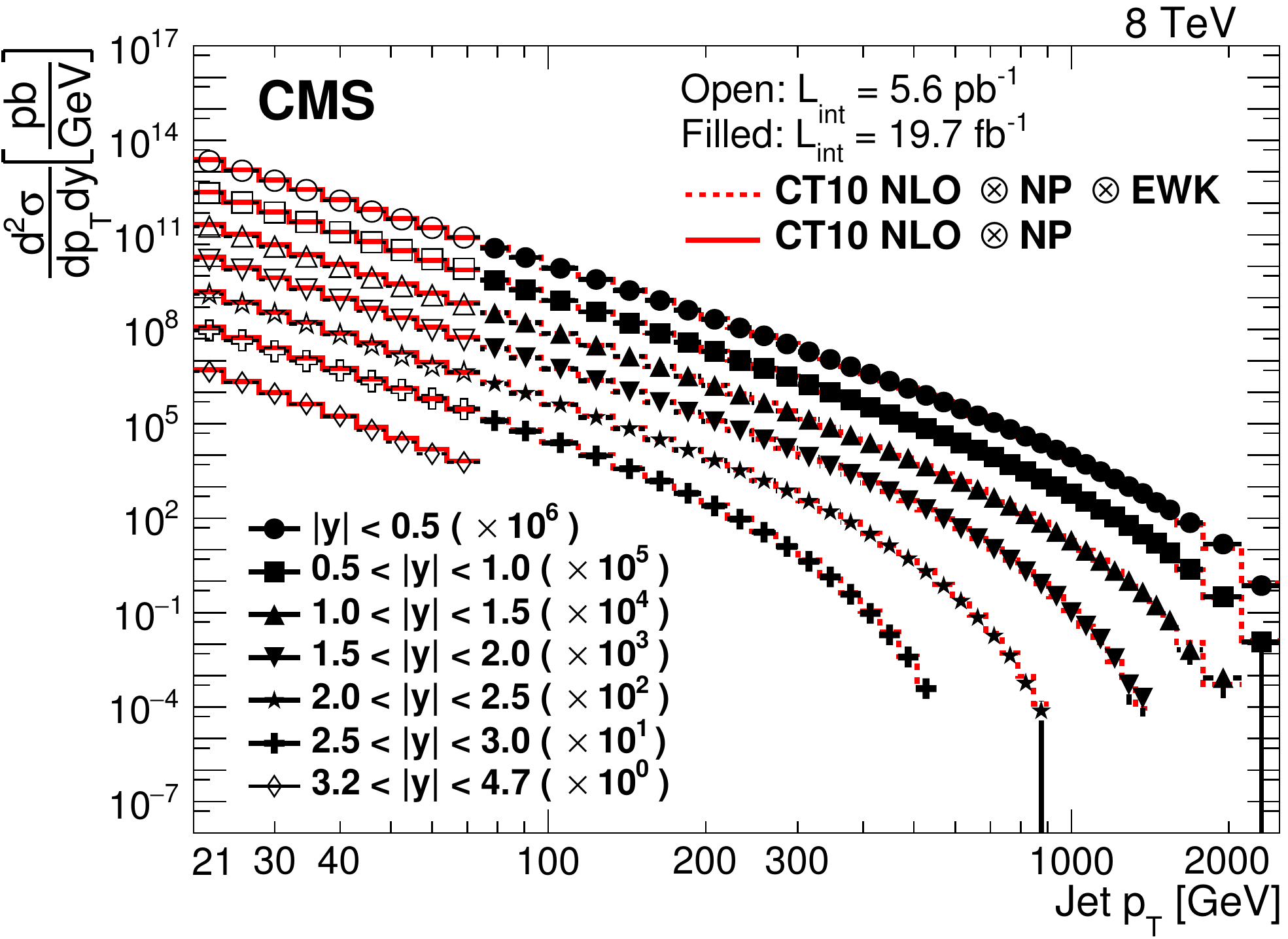}%
  \hspace{.2cm}%
  \includegraphics[width=0.5\linewidth]{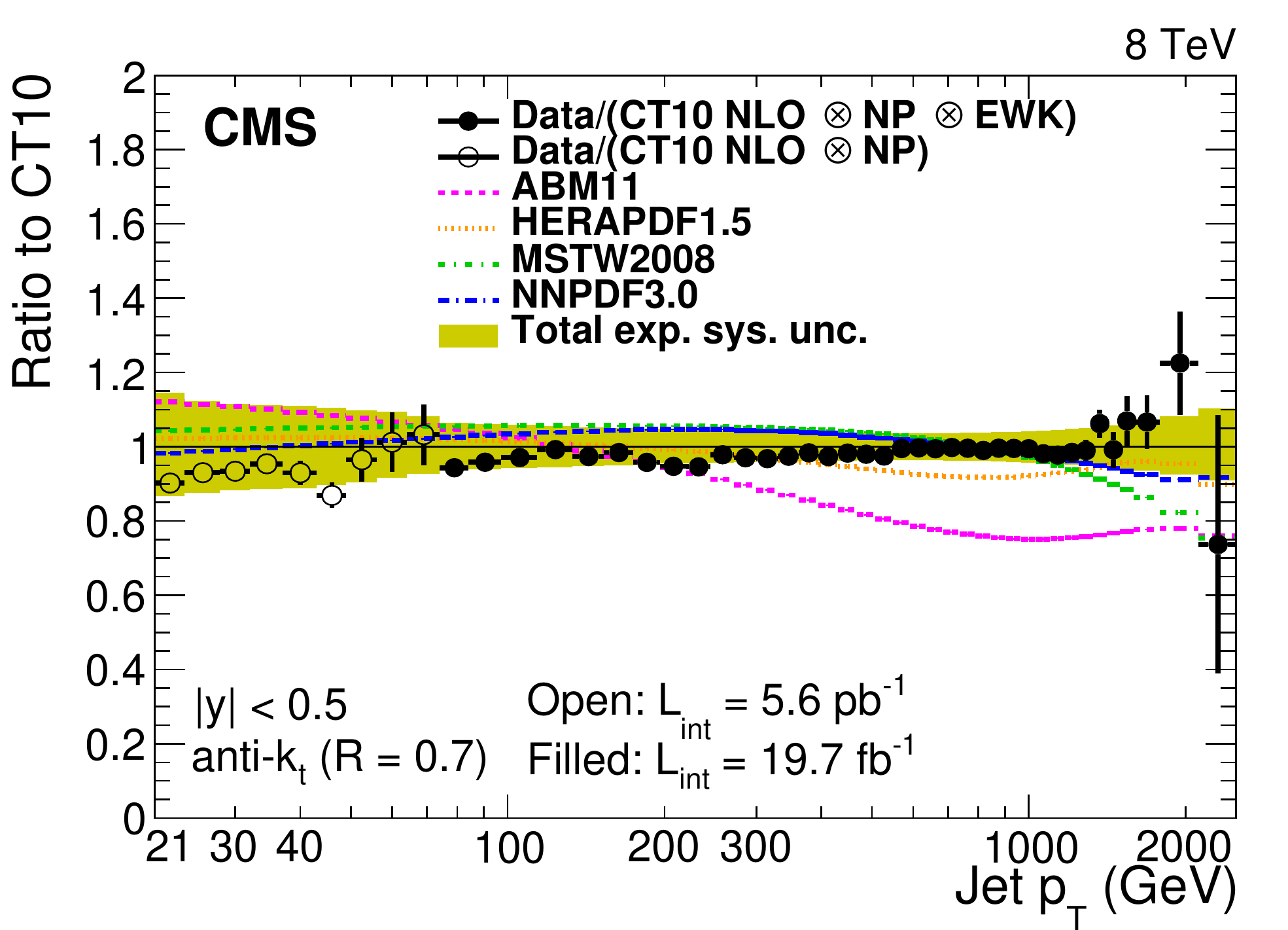}
  \caption{Double-differential inclusive jet cross-section measurement
    at $\sqrt{s}=8\TeV$ from Ref.~\cite{Khachatryan:2016wdh}. (left)
    gives an overview of all data points compared to the NLO QCD
    prediction. (right) shows the ratio of the data at central rapidity to
    NLO predictions with different PDF sets.} \label{fig:lhc:jet}
\end{figure}

In the perturbative regime, the strong interaction of quarks and
gluons leads to an abundant production of hadronic jets
with high transverse momentum. The ATLAS and
CMS collaborations have performed a series of measurements at energies
of $\sqrt{s} = 2.76$--$13\TeV$~\cite{Khachatryan:2016wdh,
  Khachatryan:2016mlc, Sirunyan:2017skj, Aaboud:2017wsi,
  Aaboud:2017dvo}. \Fref{fig:lhc:jet}(left) shows an overview of a
typical measurement of inclusive jet cross sections
double-differentially in the transverse momentum $p_T$ and rapidity
$y$ of the jet together with a QCD prediction at NLO accuracy,
supplemented by corrections for electroweak and non-perturbative
effects, where the latter include corrections of underlying event and
hadronisation. Inclusive jet measurements have the potential to cover
almost the full kinematic plane of accessible parton dynamics, as can
be appreciated in \fref{fig:lhc:xq2}(left). The data typically
agrees with the QCD predictions to $5-10\%$ as shown in
\fref{fig:lhc:jet}(right), which is truly remarkable given the wide
kinematic range and many orders of magnitude in the cross section
probed. The cross section at low $p_T$ is gigantic in comparison to
the luminosity the LHC is able to deliver and measurements are
typically performed using small data sets taken at low instantaneous
luminosity.

\begin{figure}
\centerline{\includegraphics[width=0.98\linewidth]{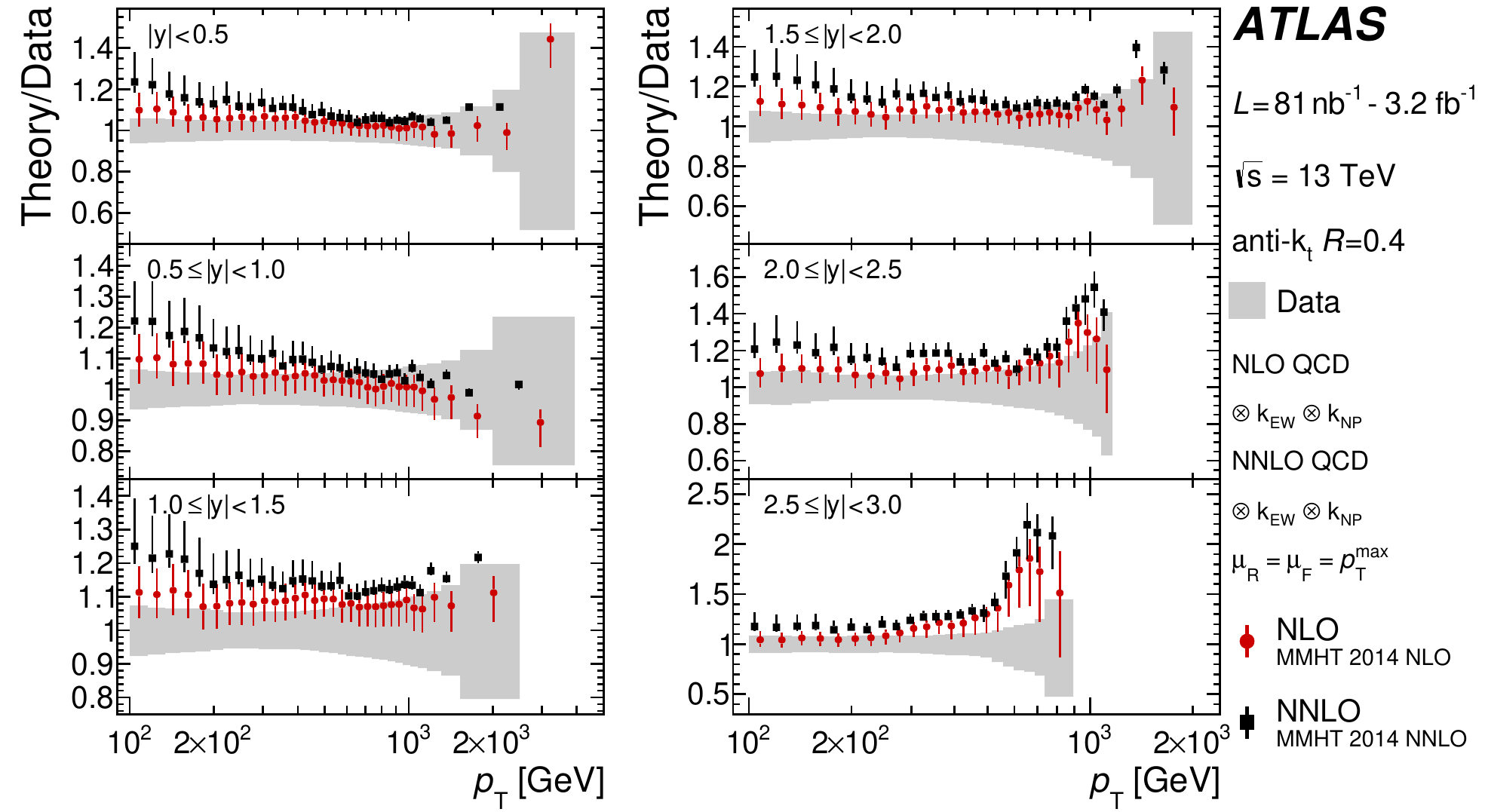}}
\caption{Double-differential inclusive jet cross-section measurement
  at $\sqrt{s}=13\TeV$ from Ref.~\cite{Aaboud:2017wsi} compared
  to NLO and NNLO QCD predictions with the scale set by the leading
  jet $p_T$.}
\label{fig:lhc:jetnnlo}
\end{figure}

The measurements at higher $p_T>100\GeV$, where the effects of
non-perturbative corrections are progressively less important, may
provide information on the gluon PDF at larger $x$ and are routinely
used in PDF fits. The recent completion of calculation of the NNLO
corrections~\cite{Currie:2016bfm, Currie:2017ctp} is a milestone to
put the interpretation of the data on a more solid footing and
promises a reduction of the theoretical uncertainties to a level less
than the PDF and the measurement uncertainties. However, contrary to
the common expectation, a significant uncertainty from scale
variations remains at NNLO. In addition, the choice of the scale as
either the leading jet $p_T$ or the individual jet $p_T$ has a
significant impact on the visual agreement with the data, where the
latter choice even leads to a worse agreement with the data than the
NLO prediction. Both of these features can be appreciated in
\fref{fig:lhc:jetnnlo} in comparison to a recent measurement. Future
studies of the theory calculation and definition of the experimental
measurement may be able to resolve this issue in the interpretation of
the data.

An interesting extension to inclusive hadronic jets are measurements
of jets produced in the fragmentation of heavy quarks. The production
of charm and beauty hadrons has been performed down to very low
transverse momentum and in the forward region by the LHCb
collaboration~\cite{Aaij:2015bpa}. This data is sensitive to the gluon
density at very low $x \sim 10^{-6}$, a region not probed by any other
collider data so far. However, the theoretical uncertainties on the NLO
predictions are large, up to $\sim 30\%$ and the interpretation relies
on the cancellation of these large uncertainties in suitably defined
ratios~\cite{Zenaiev:2015rfa, Gauld:2016kpd}.

\begin{figure}[th]
  \includegraphics[width=0.49\linewidth]{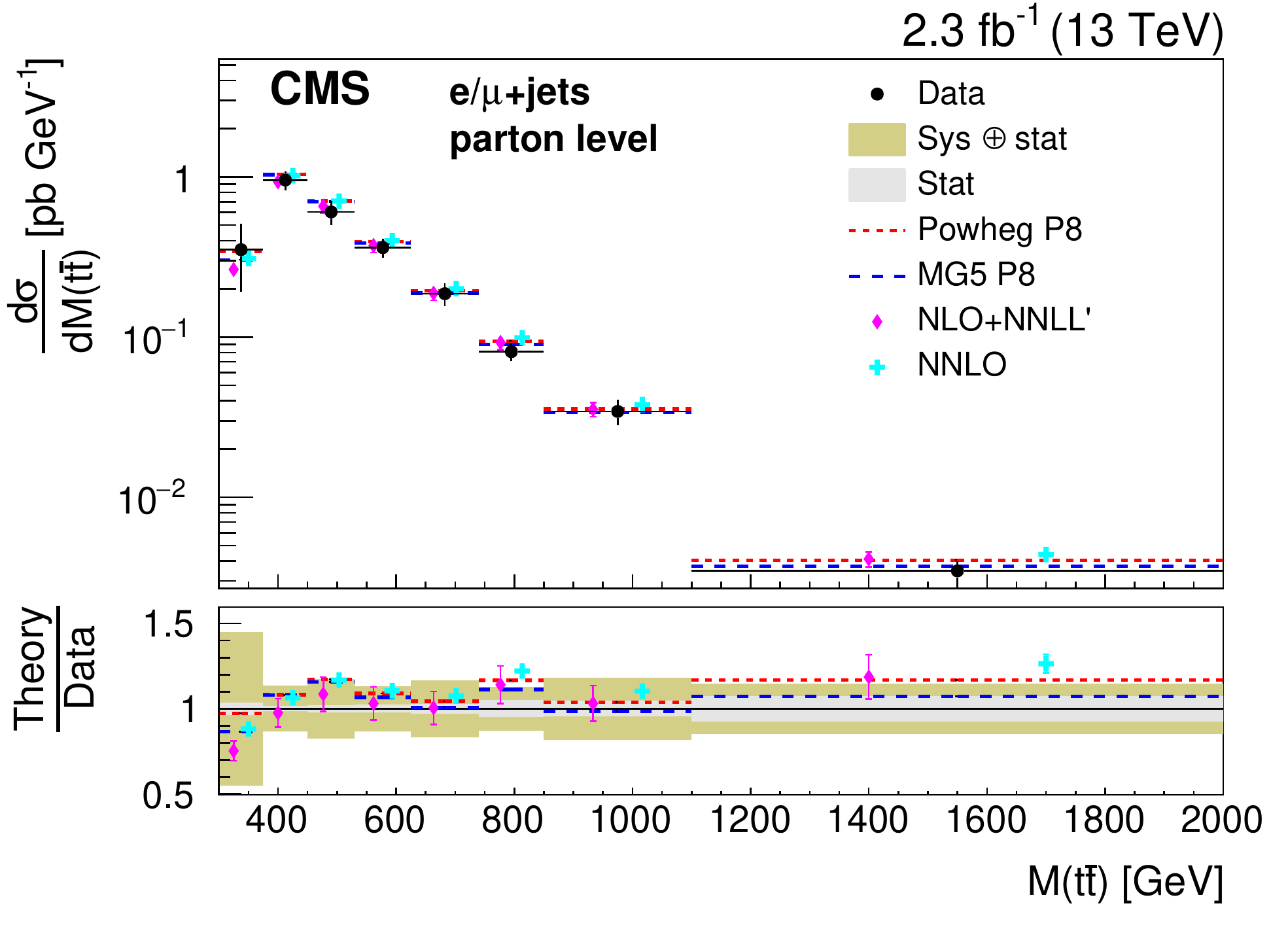}%
  \hspace{.2cm}%
  \includegraphics[width=0.49\linewidth]{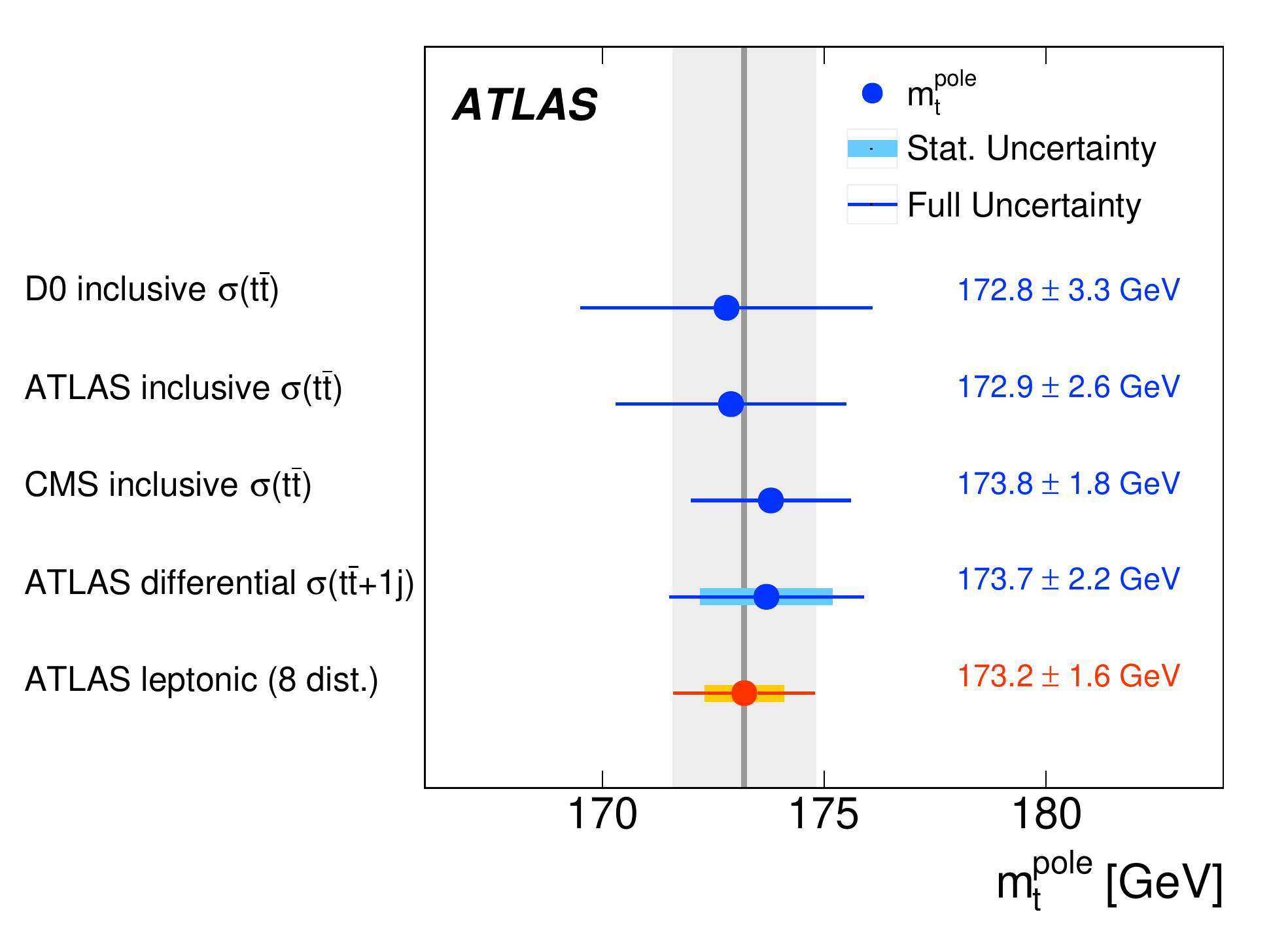}
  \caption{Example of recent measurements with top-quarks: (left) shows a
measurement of the differential cross section for $t\bar{t}$
production at $\sqrt{s}=13\TeV$ compared to a selection of predictions
by MC programs and fixed-order
    calculations~\cite{Khachatryan:2016mnb}. (right) shows a recent
    compilation of top-quark mass extractions using cross-section
    data~\cite{Aaboud:2017ujq}} \label{fig:lhc:top}
\end{figure}

The LHC has moved the study of top-quark pair production through the
strong interaction to a new level compared to the TeVatron, where the
top-quark was discovered and many studies were pioneered. The
remainder of this chapter will not be able to do justice to this
active field, but just give a glimpse of the more recent
results. Predictions for top-quark pair production are available up to
NNLO QCD differentially in the stable top-quark
kinematics~\cite{Czakon:2015owf} and these are generally seen to
describe the data well, as shown in \Fref{fig:lhc:top}(left) for a recent
measurement differential in the $t\bar{t}$-system mass. The
measurement of the top-quark mass has reached a precision of
about $0.5\GeV$~\cite{Khachatryan:2015hba, ATLAS-CONF-2017-071}. At
this level it is likely that measurements of
the ``MC mass'' differ from the (theoretically desired)
``pole mass'' for example due to
non-perturbative and colour-reconnection effects. Methods exploiting
the total~\cite{Khachatryan:2016mqs} or differential cross-section
measurements~\cite{Aaboud:2017ujq} are better
controlled with respect to non-perturbative effects and have recently
reached uncertainties of well below $2\GeV$ as shown in
\Fref{fig:lhc:top}(right), typically limited by the uncertainties on the
PDFs and missing higher-order corrections.

Finally, the production of jets may be used to probe the 
strong coupling \alphas, which is obviously ``a crucial ingredient of
perturbative QCD''~\cite{Altarelli:2013bpa} and in fact
the least-well known of the fundamental couplings with a
precision of about $1\%$ at a scale of the $Z$-boson
mass. Unfortunately it is not possible to extract
$\alphas(m_Z)$ from LHC jet measurements with competitive
precision. Instead, the data may be used to probe \alphas\ at very
large scales of up to
$2\TeV$~\cite{Khachatryan:2016mlc}. Determinations from the $t\bar{t}$
cross section have been shown to reach an improved precision compared
to inclusive jets~\cite{Chatrchyan:2013haa, Klijnsma:2017eqp}.
The uncertainty in \alphas\ is important for several LHC
predictions, such as Higgs-boson production. The size of the
uncertainty in \alphas\ is a topic of debate and Guido Altarelli
certainly had an opinion on this issue, advocating caution to the
estimates of theory uncertainties~\cite{Altarelli:2013bpa}.

\section{Di-lepton production in the Drell--Yan Process}
\label{sec:lhc:dy}

Di-lepton production in hadron collisions through the Drell--Yan (DY)
process~\cite{Drell:1970wh} has played a fundamental role in
establishing the SM. It is here considered to include both
the neutral current $q\bar{q} \to Z/\gamma^* \to \ell^+\ell^-$ as well
as the charged current $q\bar{q}' \to W^\pm \to \ell^\pm
\overset{\scriptscriptstyle(-)}{\nu}$ reactions. Of critical importance was the
discovery of the heavy intermediate vector bosons $W$ and $Z$ by the
UA1 and UA2 experiments~\cite{Arnison:1983rp, Banner:1983jy,
  Arnison:1983mk, Bagnaia:1983zx} in DY reactions. Beyond these
discoveries, the DY process was a test-bed to develop the procedure to
systematically calculate and improve cross sections in
hadron-collisions through the factorization theorem and an expansion
in orders of \alphas. In fact, this process remains one
of the few examples in hadron--hadron collisions, where factorization
has been rigorously proven to
hold~\cite{Collins:1983ju,Bodwin:1984hc}.

Each major step forward in the theory calculation typically took a
decade. The first step beyond the initial leading order calculation came through
major contributions of the ``Rome group'' with Guido
Altarelli~\cite{Altarelli:2011zv}, who completed ``one of the first
calculations of NLO corrections in QCD'' in
1978/79~\cite{Altarelli:1978id,Altarelli:1979ub,KubarAndre:1978uy,Kubar:1980zv}.
A critical part was to treat the PDFs and their evolution in a
consistent way in DIS and DY reactions. The NLO corrections ``turned
out to be surprisingly large''~\cite{Altarelli:2011zv}, necessitating
calculations beyond NLO. The NNLO corrections were completed in the
early 1990s~\cite{Hamberg:1990np, vanNeerven:1991gh, Rijken:1994sh}
and the fully differential NNLO predictions became available until
about 2010~\cite{Anastasiou:2003ds, Melnikov:2006kv, Catani:2009sm,
  Gavin:2010az}.

\begin{figure}
  \centerline{\includegraphics[width=0.65\linewidth]{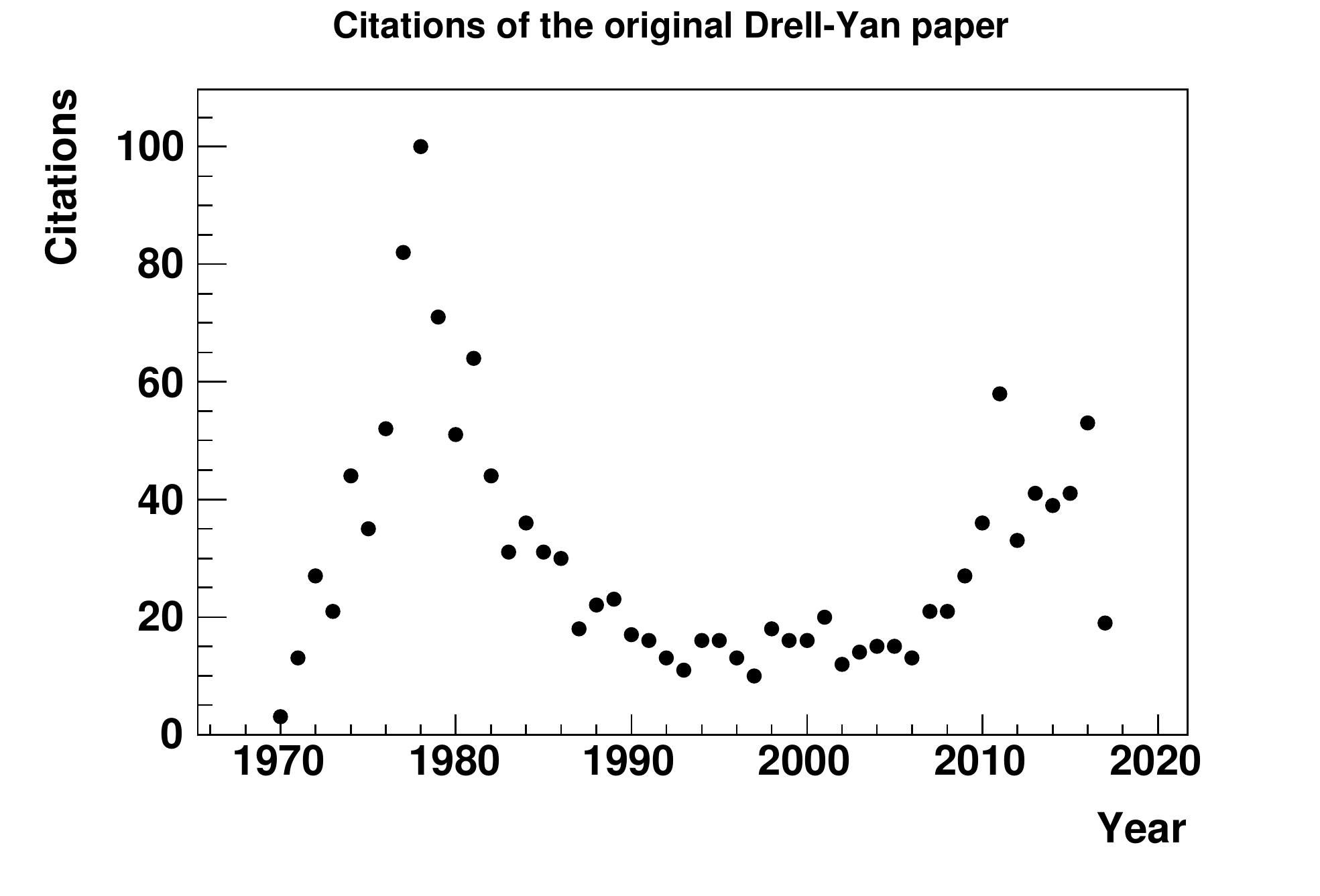}}
  \caption{History of citations of the original paper on the
    Drell--Yan process~\cite{Drell:1970wh} taken from \texttt{inspire.net}.}
  \label{fig:lhc:dycitations}
\end{figure}

\Fref{fig:lhc:dycitations} shows the number of citations of the
original paper by Drell and Yan as a function of time. This reflects
the initial interest in the process to establish the SM in the 1980s,
as well as the more recent ``resurrection'' that is closely linked to
new analyses of this process at the LHC. This renewed interest has
diverse reasons, which includes the study of PDFs, precision
measurements of SM parameters as the $W$-boson mass and the
effective weak mixing angle $\sin^2\theta_W$, and the background to
searches beyond the SM as for example particle candidates for dark
matter~\cite{Lindert:2017olm}. The measurements are facilitated by
large samples of tens of millions of events reconstructed with small
backgrounds and well-understood detector performance for the
final-state leptons.

\begin{figure}[th]
  \includegraphics[width=0.455\linewidth]{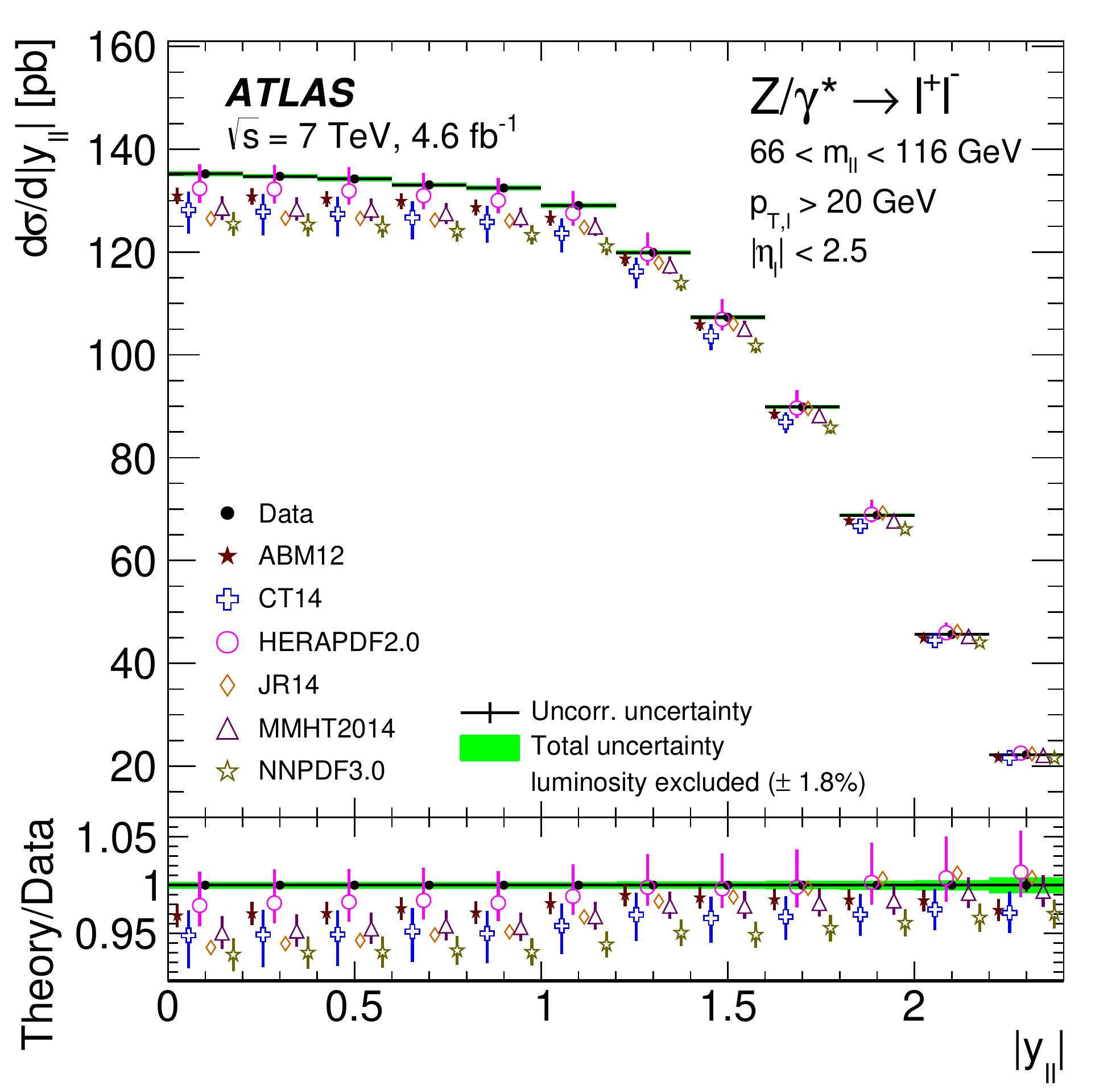}%
  \hspace{.2cm}%
  \includegraphics[width=0.53\linewidth]{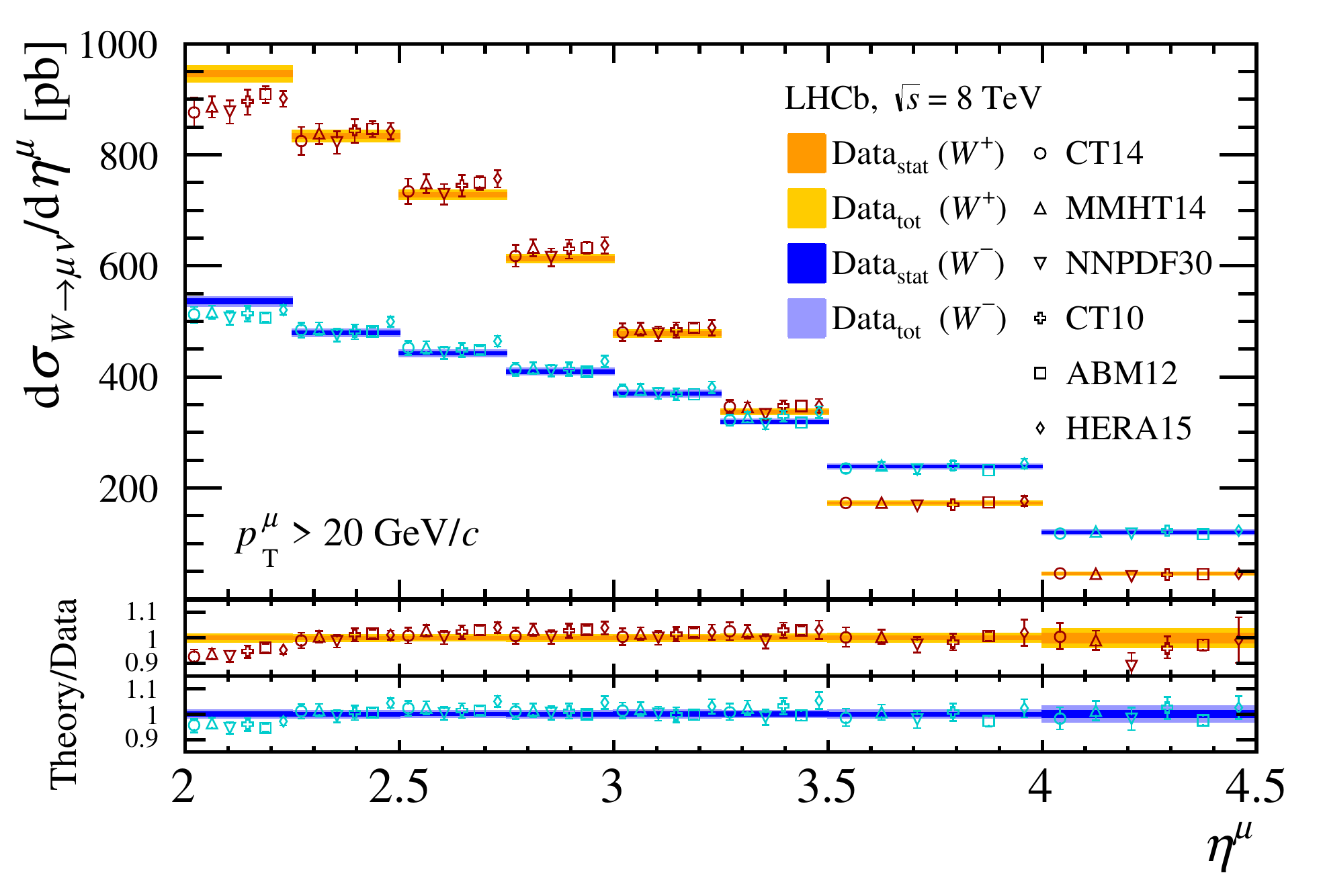}
  \caption{Two examples of recent Drell--Yan measurements at the LHC
    compared to NNLO QCD predictions with a variety of different PDF
    sets. (left) shows a measurement of the differential $Z/\gamma^*$
    cross section as a function of the di-lepton rapidity
    $y_{\ell\ell}$ at $\sqrt{s}=7\TeV$ \cite{Aaboud:2016btc} (right) shows
    a measurement of $W^+$ and $W^-$ production as a function of the
    lepton pseudo-rapidity $\eta_\ell$~
    \cite{Aaij:2015zlq}} \label{fig:lhc:wzxsec}
\end{figure}

Measurements of differential cross sections and distributions for
$Z/\gamma^*$ and $W^\pm$ production have been performed by the ATLAS,
CMS and LHCb collaboration in $pp$ collisions at $\sqrt{s}
=2.76$--$13\TeV$ and as a function of many kinematic variables. The
classical variables to constrain PDFs are the di-lepton rapidity and
mass, $y_{\ell\ell}$ and $m_{\ell\ell}$ for $Z/\gamma^*$ and the
lepton rapidity $\eta_\ell$ for $W^\pm$ production with some recent
examples given by
Refs.~\cite{Aaboud:2016btc,Aad:2016zzw,CMS:2014jea,Khachatryan:2016pev,
  Aaij:2015zlq,Aaij:2016mgv}. \Fref{fig:lhc:wzxsec} shows two examples
of such measurements compared to state-of-the-art cross-section
calculations with different PDF sets. The measurements have reached an
accuracy of a percent or better, which is below the uncertainties
of a typical PDF set and often also below the uncertainty of the NNLO
theory calculation. The fact that these precise $W^\pm$ and
$Z/\gamma^*$ cross-section data generally agree with the predictions, is another
testament to the success of our understanding of perturbative
QCD.

\begin{figure}[th]
  \includegraphics[width=0.48\linewidth]{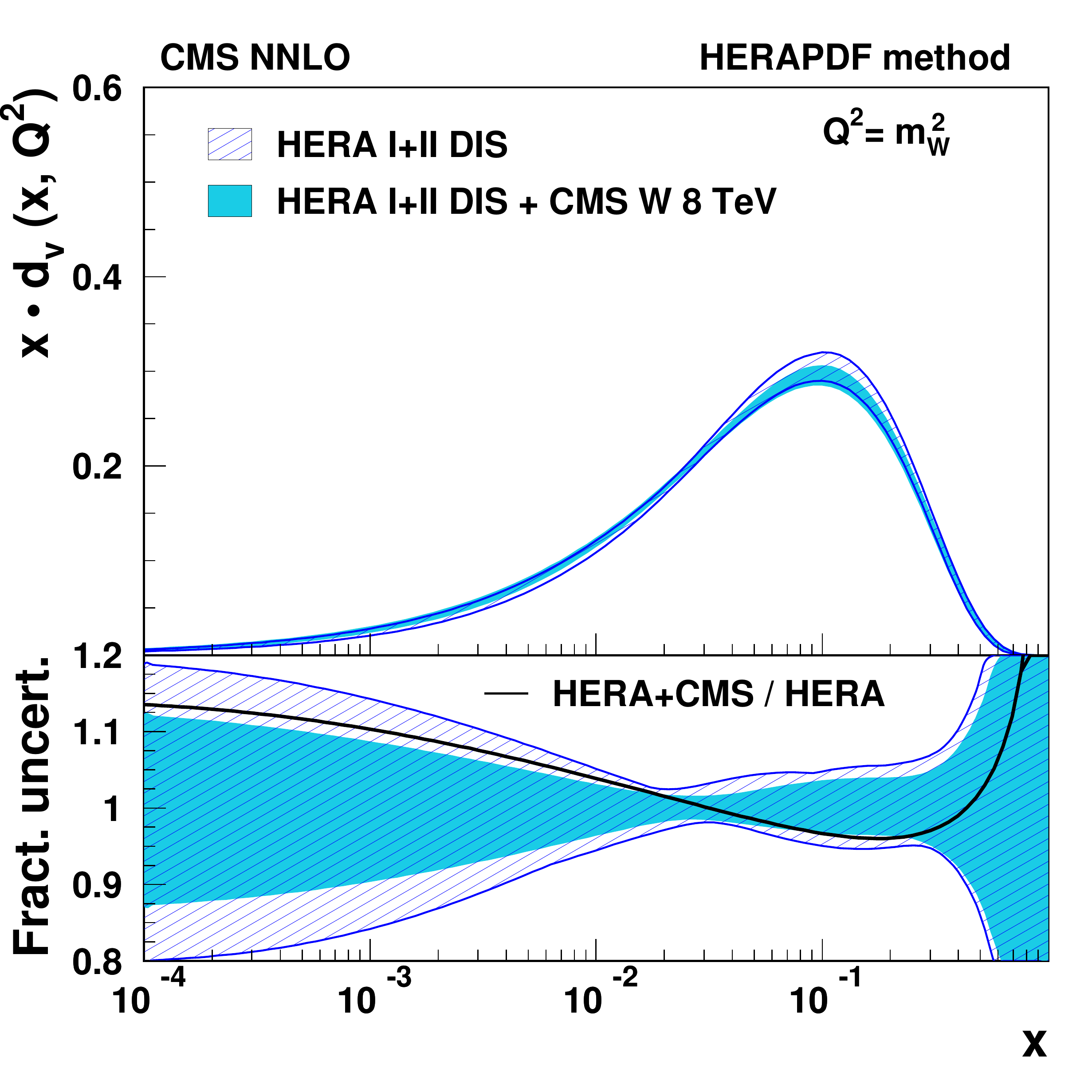}
  \includegraphics[width=0.52\linewidth]{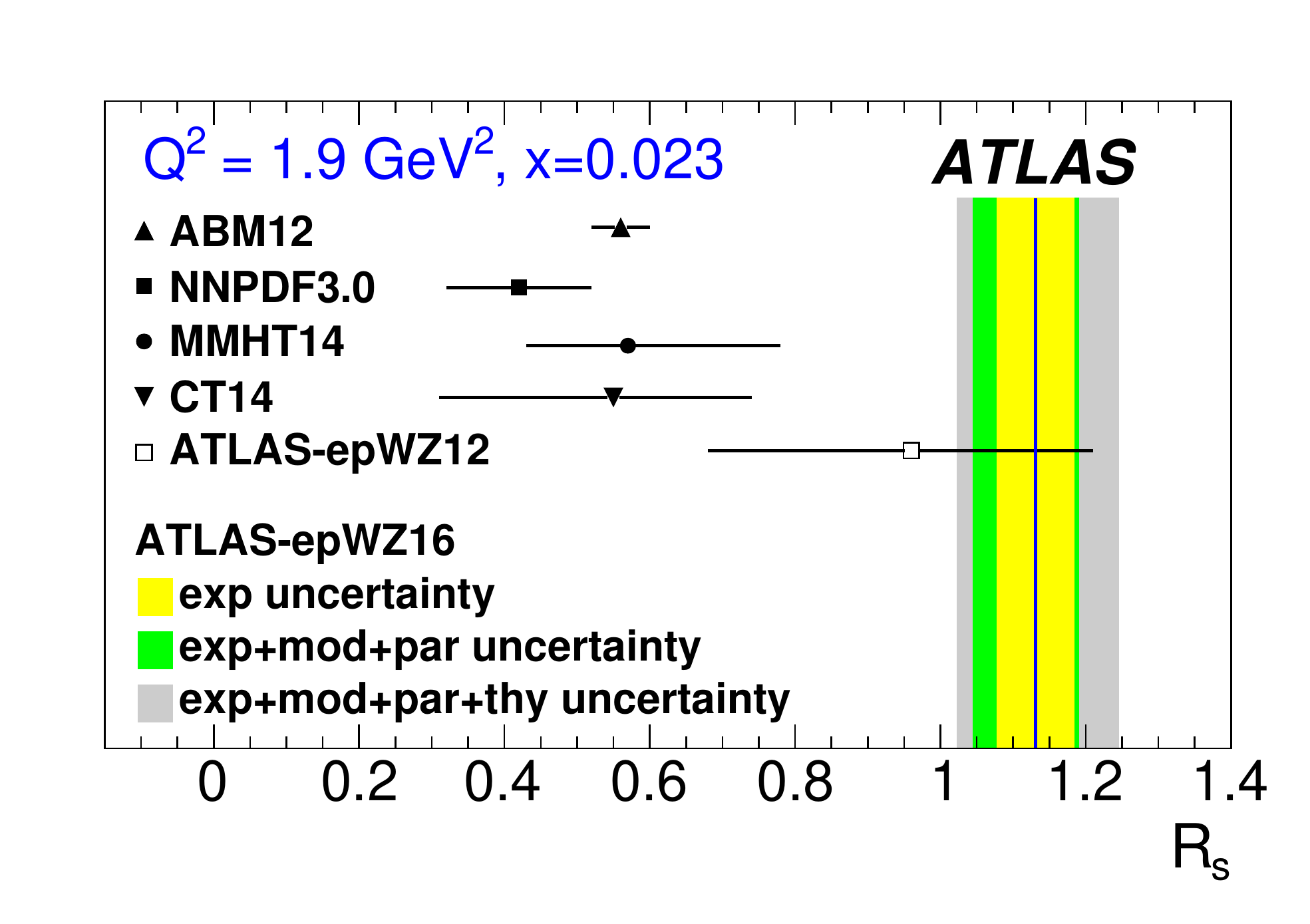}
  \caption{Two examples showing the impact of LHC $W^\pm$ and
    $Z/\gamma^*$ production data on the knowledge on the proton
    PDFs. (left) shows a study of including differential $W^\pm$ data in
    a PDF fit based on HERA DIS data, comparing the uncertainties on
    the $d$-valence density with and without the $W^\pm$
    data~\cite{Khachatryan:2016pev} (right) show the ratio $R_s =
    (s+\bar{s})/(\bar{u}+\bar{d})$ as a measure of the size
    strange-quark distribution compared to the up- and down-quark
    sea. The result from a fit with LHC $Z/\gamma^*$ and $W^\pm$ data
    labelled as ``ATLAS-epWZ16''~\cite{Aaboud:2016btc} is compared to the prediction of other
    PDF fits~\cite{Alekhin:2013nda,Ball:2014uwa,Harland-Lang:2014zoa,Dulat:2015mca,Aad:2011dm}.} \label{fig:lhc:wzpdf}
\end{figure}

Such data can now be included in PDF analyses based on
either the HERA DIS data set~\cite{Abramowicz:2015mha} or global PDF
fits~\cite{Harland-Lang:2014zoa,Dulat:2015mca,Alekhin:2017kpj,Ball:2017nwa}
to improve our knowledge on the proton structure in broadly
speaking three areas. Firstly, the LHCb data gives unique information
on PDFs at very low Bjorken $x \sim 10^{-4}$ or very large $x \sim
0.5$, a range where the information of the HERA data set is limited and
information at high $x$ from fixed-target data may not be fully reliable.
 Second, as illustrated in
\Fref{fig:lhc:wzpdf}(left), data on $W^\pm$-boson production helps to improve the information on
low and intermediate $x$ valence quarks distributions. Finally, $Z$ production
at the LHC at central rapidity has a significant contribution from strange quarks,
which are only constrained from
charm production in neutrino--nucleon DIS on heavy targets at somewhat larger
$x$. As demonstrated in \Fref{fig:lhc:wzpdf}(right), the LHC data has here
a unique impact and strongly hints at a larger strange-quark
density than previously inferred from the neutrino data.

\begin{figure}[th]
\centerline{\includegraphics[width=0.98\linewidth]{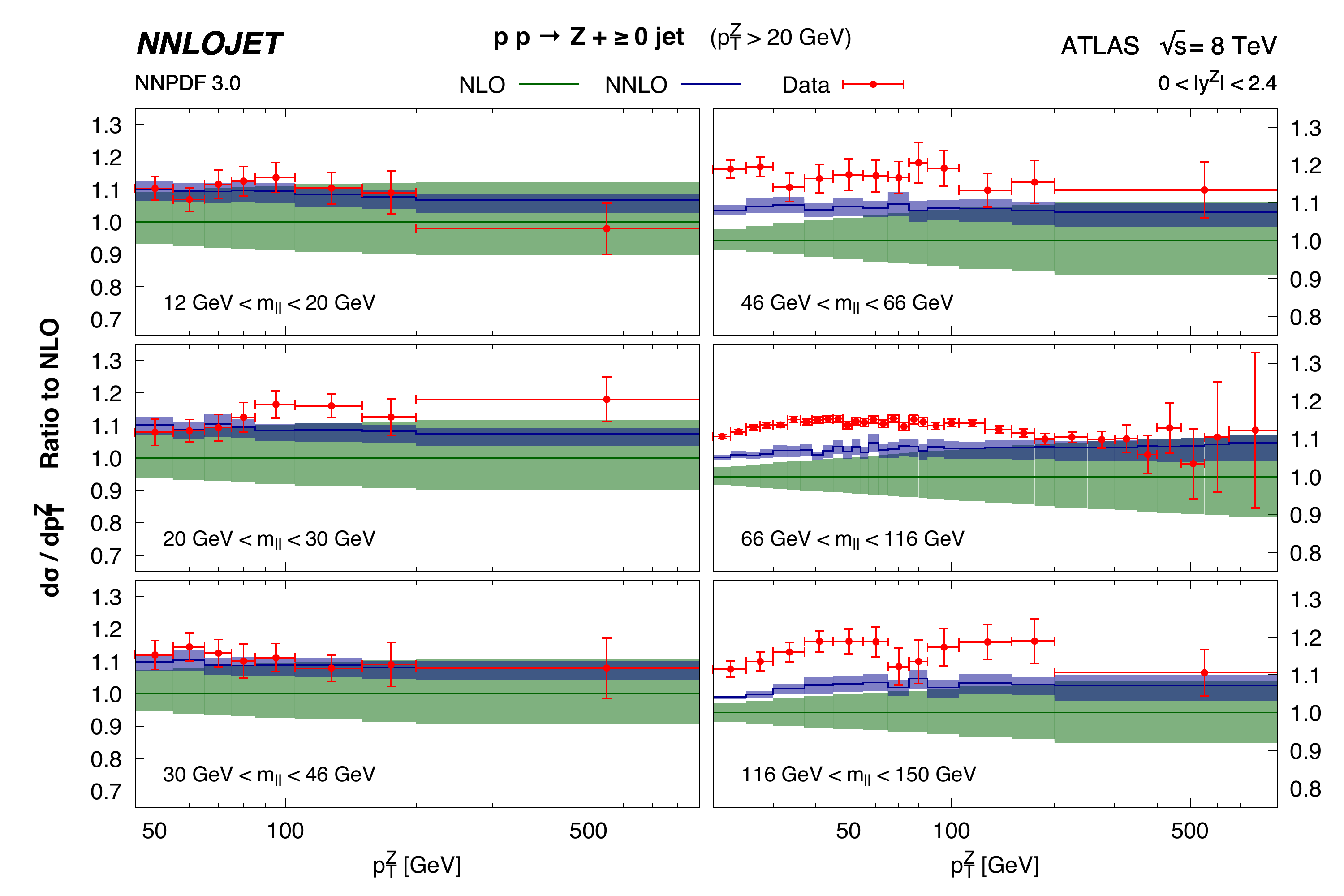}}
\caption{Double-differential di-lepton cross-section measurement as a
  function of mass and transverse momentum at $\sqrt{s}=8\TeV$
  compared to a fixed-order NNLO calculation~\cite{Ridder:2016nkl}.}
\label{fig:lhc:zpt}
\end{figure}

Another key variable is the transverse momentum of the boson, which
poses an interesting theoretical challenge and has implications for
many studies at the LHC. The distribution is more easily measured in
the neutral current process as the di-lepton $p_{T,\ell\ell}$, see
e.g. Refs.~\cite{Aad:2015auj, Khachatryan:2015oaa,
  Khachatryan:2016nbe}. The spectrum can be roughly separated into two regimes
at a transverse momentum of above and below half the boson mass. The
higher-$p_{T,\ell\ell}$ region may be described by a fixed-order
calculation, while an all-order resummation calculation is required to
describe the lower-$p_{T,\ell\ell}$ domain that is dominated by
multiple soft-gluon emission and possibly also influenced by
non-perturbative effects. \Fref{fig:lhc:zpt} shows the high
$p_{T,\ell\ell}> 20\GeV$ part of a measurement differential in both
the di-lepton transverse momentum and mass compared to a very recent
NNLO fixed-order calculation for the $pp \to Z+\mathrm{jet}$ process
at order $\alpha_S^3$~\cite{Ridder:2015dxa, Boughezal:2015ded}. The
correction to the $\alpha_S^2$ calculation is seen to be
significant, although within the uncertainty estimate in the range
$p_{T,\ell\ell} \gtrsim m_{\ell\ell}$. In parts of the phase space,
especially near the $Z$-boson resonance, the data is more precise even
than the $\alpha_S^3$ prediction. There remains a tension between the
data and the prediction that may be related to the PDFs and can possibly
be improved by including the data in a PDF analysis as discussed in
Ref.~\cite{Boughezal:2017nla}.

The large data sets of $Z/\gamma^*$ and $W^\pm$ can be used for
precision measurements of the effective weak mixing angle
$\sin^2\theta_W$ and the $W$-boson mass, respectively. These
measurements are interesting to pin down the fit of precision
electroweak observables~\cite{Baak:2014ora, deBlas:2016ojx} and search
indirectly for deviations from the SM~\cite{Heinemeyer:2013dia}. The
$\sin^2\theta_W$ measurement exploits the change of the
``Forward-Backward Asymmetry'' in the $q\bar{q} \to Z/\gamma^* \to
\ell^-\ell^+$ process in the vicinity of the $Z$ resonance. The
$W$-boson mass can be extracted through minuscule changes in the
shapes of kinematic variables such as the lepton $p_T$ in $W$ events.

These electroweak measurements require an extreme experimental
precision, for example the lepton momentum calibration has to be
understood to the $10^{-4}$ level. Even more critical is the control
of uncertainties related to the modelling of the sensitive
distributions with respect to QCD effects related to PDFs,
heavy-flavour initiated processes and the transverse-momentum spectrum
of the bosons. The LHC environment poses such a challenge, that Guido
Altarelli as many of his peers was skeptical about the feasibility of
such an enterprise, stating that ``... anyone [who] wants to measure
the $W$ mass at the LHC with high precision, better he commits suicide
...''~\cite{guidolhcc98}.

\begin{figure}[th]
  \includegraphics[width=0.49\linewidth]{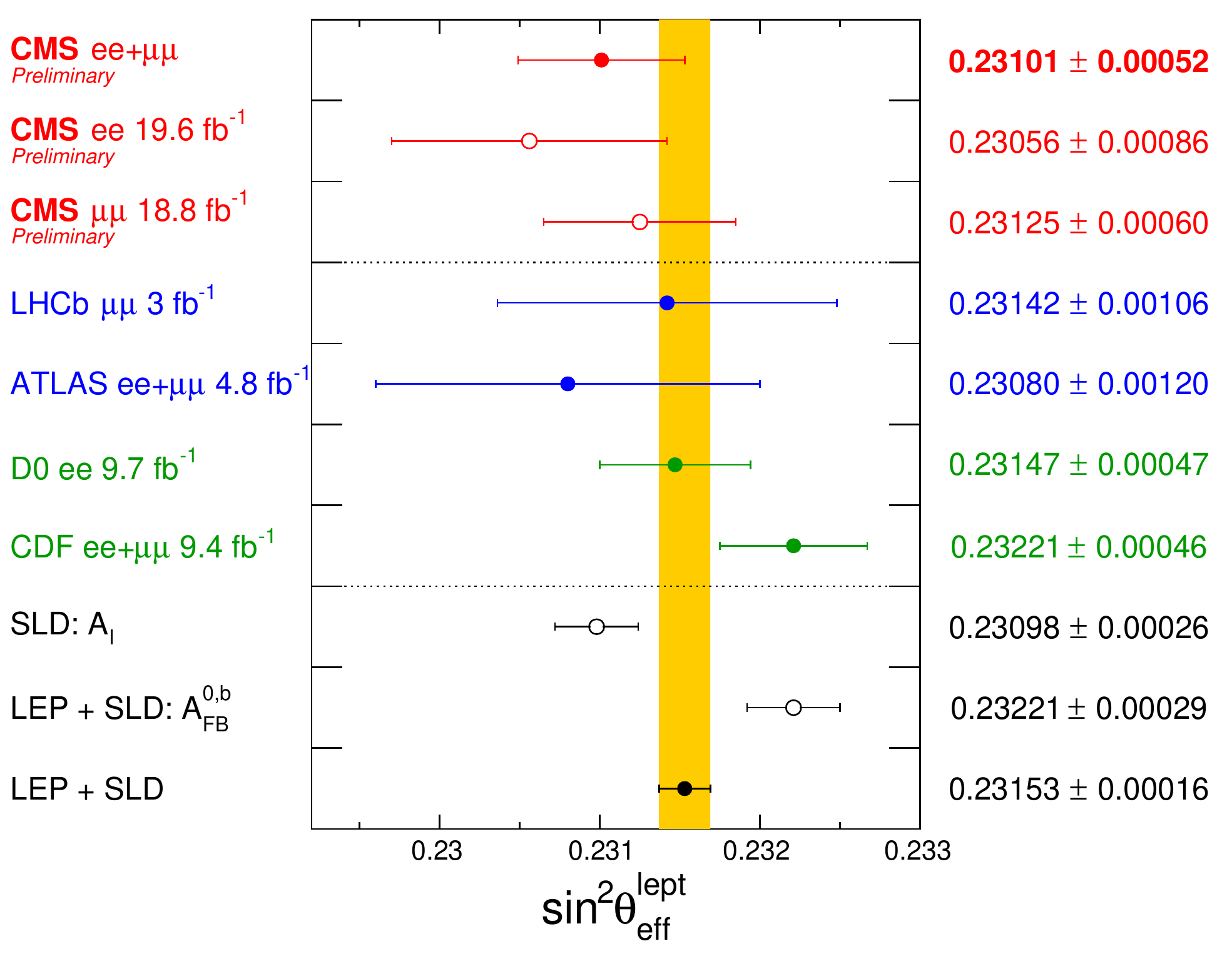}%
  \hspace{.2cm}%
  \includegraphics[width=0.49\linewidth]{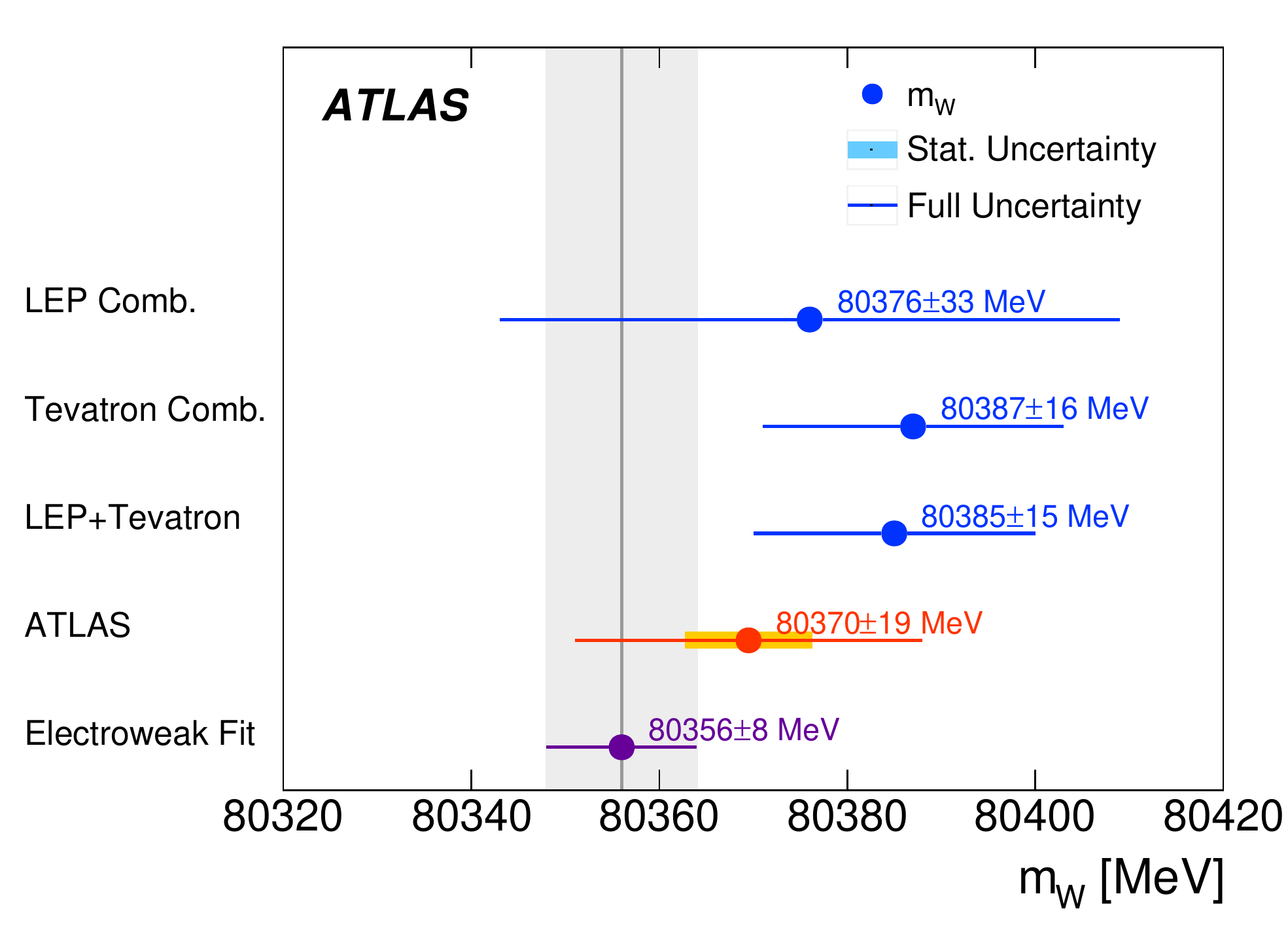}
  \caption{Overview of measurements of precision electroweak
    parameters at different colliders for (left) the effective weak
    mixing angle $\sin^2\theta_W$~\cite{CMS-PAS-SMP-16-007} and (right)
    the mass of the $W$ boson~\cite{Aaboud:2017svj}. }
  \label{fig:lhc:sin2thmw}
\end{figure}

The achievements of the first competitive measurements of the
$W$-boson mass~\cite{Aaboud:2017svj} and
$\sin^2\theta_W$~\cite{CMS-PAS-SMP-16-007} in spite of the major
effort are therefore all the more
remarkable. \Fref{fig:lhc:sin2thmw}(left) gives an overview of
$\sin^2\theta_W$ from $e^+e^-$ colliders, the TeVatron and the LHC.
The most recent CMS measurement with the $\sqrt{s}=8\TeV$ data set
reaches a precision similar to the TeVatron experiments and
``just'' a factor two worse than the LEP and SLD results. In
\Fref{fig:lhc:sin2thmw}(right) an overview of measurements of the
$W$-boson mass is shown including the recent ATLAS determination using
the $\sqrt{s}=7\TeV$ data set, which has reached the same uncertainty
as the previously best measurement by CDF.

\section{Multiboson production and electroweak processes}
\label{sec:lhc:multiv}

In addition to the study of the simpler and abundant processes, the
LHC allows also the measurements of rarer and more complex final
states. In this section we will discuss processes that go beyond the
leading order in the electroweak couplings, where either two or three
bosons $V = W, Z, \gamma$ are produced or two of these bosons ``fuse''
(vector boson fusion, VBF) or ``scatter'' (vector boson scattering,
VBS). The diagrams involved contain vertices sensitive to the triple
and quartic couplings of the electroweak bosons, which are non-trivial
in the SM due to the non-Abelian $U(1)_Y \times SU(2)_L$ structure,
which is ``unbroken in the interaction vertices, ... [but] is instead
badly broken in the masses''~\cite{Altarelli:2011vt} by the Higgs
mechanism. In the SM the interplay of various diagrams, such as those
containing $\gamma WW$ and $ZWW$ vertices, guarantees a result that
does not violate unitarity. Another well-known example is the
scattering of longitudinally polarized massive vector bosons,
e.g. $W_LW_L \to W_LW_L$, where unitarity is violated already at about
$1\TeV$ without the contribution from the Higgs
boson~\cite{Veltman:1976rt, Lee:1977yc, Lee:1977eg}. Any deviation
from the SM predictions would imply physics with new forces or
particles.~\footnote{Similarly, it is interesting to study the
  electroweak properties of the top-quark, which may have a special
  role in the process of electroweak symmetry breaking due to its
  large mass. Such studies can be performed e.g. through electroweak
  production of single top-quarks or the production of top-quarks in
  association with electroweak bosons $t\bar{t}V$ or $tV$, but a
  detailed discussion is beyond the scope of this section.}

\begin{figure}
  \hspace{-1cm}%
  \includegraphics[height=0.46\linewidth]{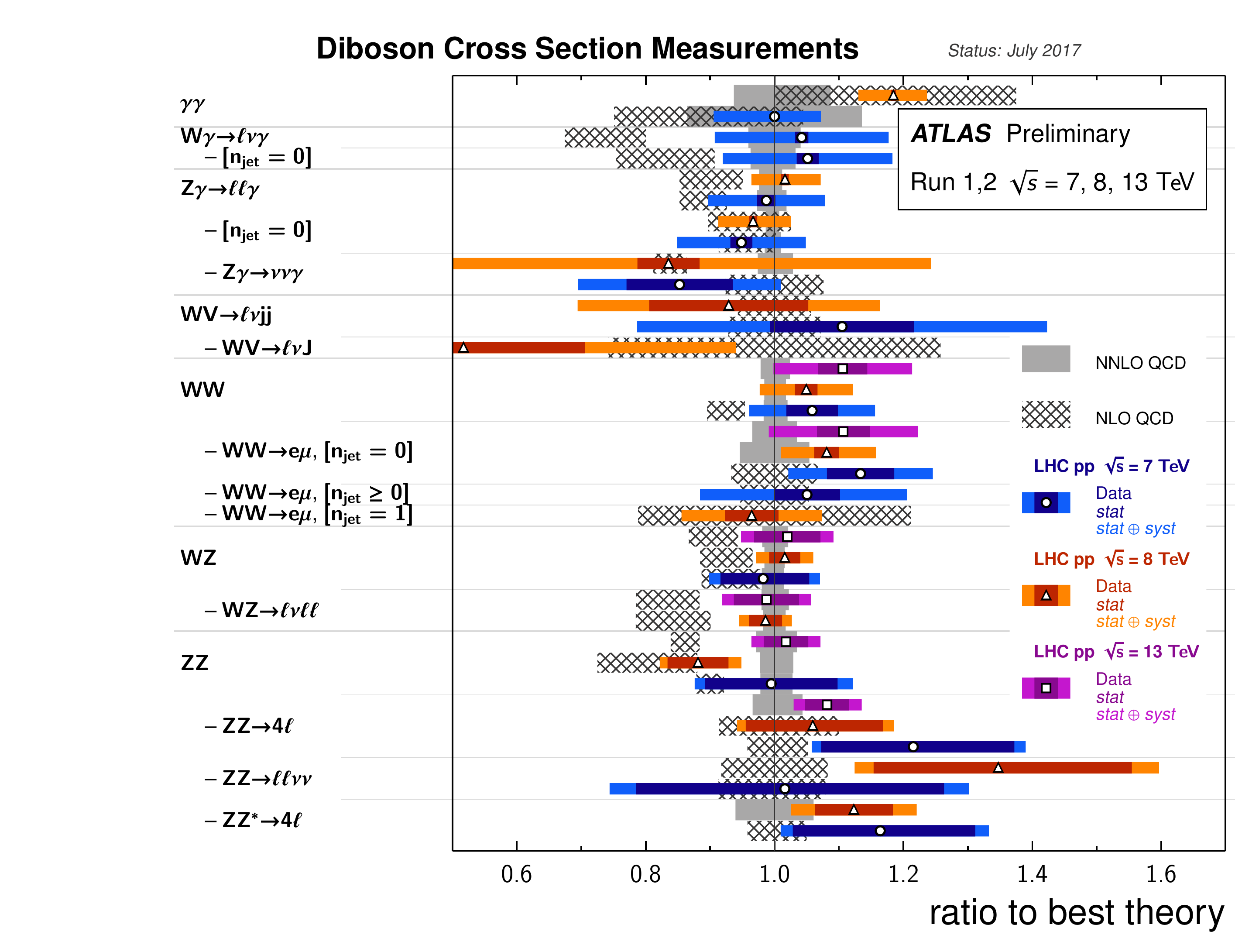}%
  \hspace{.5cm}%
  \includegraphics[height=0.45\linewidth]{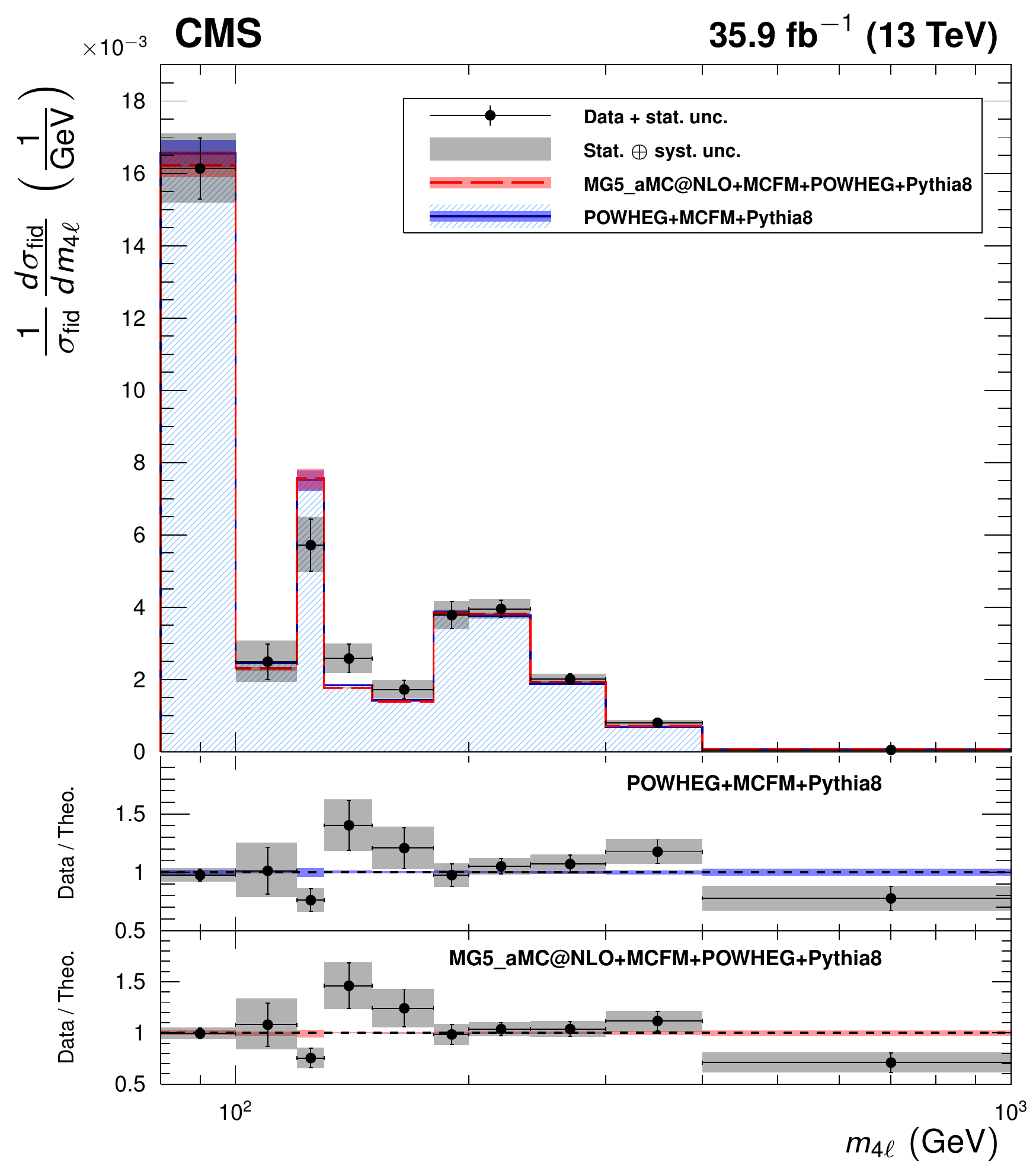}
  \caption{(left) A summary of diboson cross-section measurements by the
    ATLAS collaboration~\cite{summaryATLAS}. Measurements are compared
    to SM theory predictions calculated at NLO (hatched) and NNLO QCD
    (solid, where available). (right) The four-lepton invariant mass
    spectrum, $m_{4\ell}$, compared to the SM prediction including the
    processes $Z\to 4\ell$, $H \to 4\ell$ and $ZZ^{(*)} \to
    4\ell$~\cite{Sirunyan:2017zjc}.}
  \label{fig:lhc:multibosonoverview}
\end{figure}

In \Fref{fig:lhc:multibosonoverview}(left) an overview of measurements of
di-boson cross sections is given. All possible combinations of $VV$
processes have been measured, with the majority of measurements
considering the decays with (charged) leptons of the massive $W$ or $Z$
bosons. The precision has reached a remarkable level of about $5\%$ for
several of the combinations. An interesting feature, visible
even on this coarse overview, are the large differences between the
theory calculations at NLO and NNLO QCD, where the NNLO predictions
are in much better agreement with the data. Such features were noticed
already in the early analyses of the $W\gamma$~\cite{Aad:2013izg}, and
later explained by large NNLO
corrections~\cite{Grazzini:2015nwa}. A similar feature is also noticeable in
the $WZ$ final state~\cite{Grazzini:2017ckn}, which shares with the
$W\gamma$ process the feature of a ``radiation-zero'' suppressed LO
prediction with large higher-order QCD corrections.

For neutral combinations such as $\gamma\gamma$, $WW$ and $ZZ$, the
higher-order QCD corrections have also been found to be large, mainly
because of the gluon-initiated contribution $gg \to VV$ through a box
diagram, which appear first as part of the NNLO ($\alpha_S^2$)
calculation and receive large $\alpha_S^3$
corrections~\cite{Caola:2015psa}. The final state with four charged
leptons, $4\ell$ ($\ell = e,\mu$), is particularly instructive, as the
excellent lepton identification capabilities of the LHC detectors
allow a full reconstruction without significant backgrounds in an wide
range of four-lepton invariant masses from about $m_{4\ell} =
80$--$1000\GeV$~\cite{Aad:2015rka,Sirunyan:2017zjc}, although the rate
is low because of the small branching of the $Z$ boson to two charged
leptons. As illustrated in \Fref{fig:lhc:multibosonoverview}(right), this
distribution contains a host of interesting features: a resonance at
the mass of the $Z$ boson from the radiative decay $Z \to 4\ell$, a
resonance at the mass of the Higgs boson from its golden decay $H \to
ZZ^* \to 4\ell$, a threshold at twice the $Z$ boson mass for the
production of two on-shell $Z$ bosons, and finally a negative
interference of gluon-induced $ZZ$-production diagrams with and
without the Higgs boson, important especially above
$m_{4\ell}>2m_\mathrm{top}$, which is also sensitive to the width of
the Higgs boson~\cite{Caola:2013yja}.

In contrast to the measurements of di-boson final states, which is
entering the precision phase, the searches for tri-boson processes
will require more luminosity, with the exception of the
$W\gamma\gamma$ and $Z\gamma\gamma$ final states, for which evidence
and observations have been
reported, respectively~\cite{Aad:2015uqa,Aad:2016sau,Sirunyan:2017lvq}.

\begin{figure}
  \includegraphics[width=0.51\linewidth]{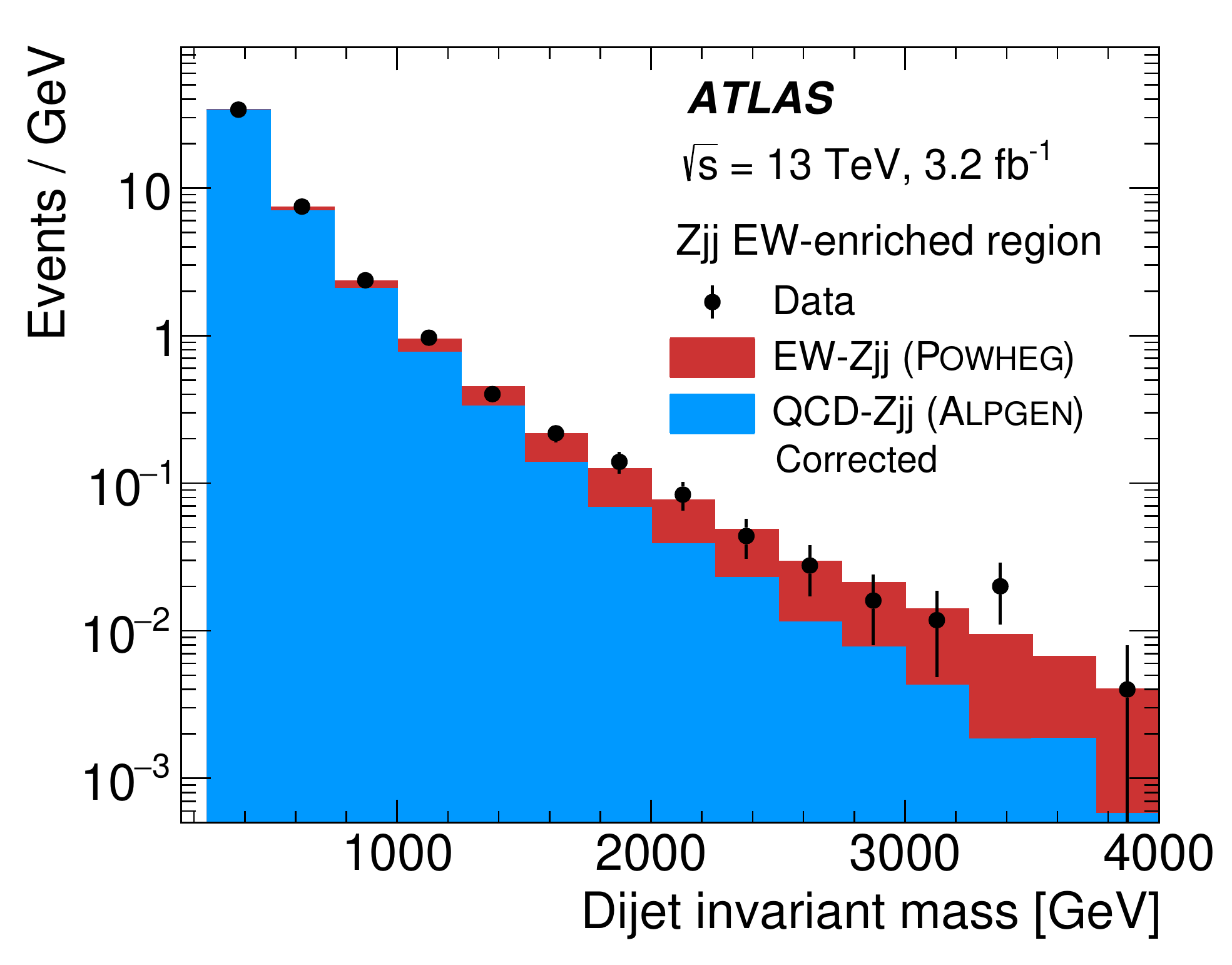}
  \includegraphics[width=0.49\linewidth]{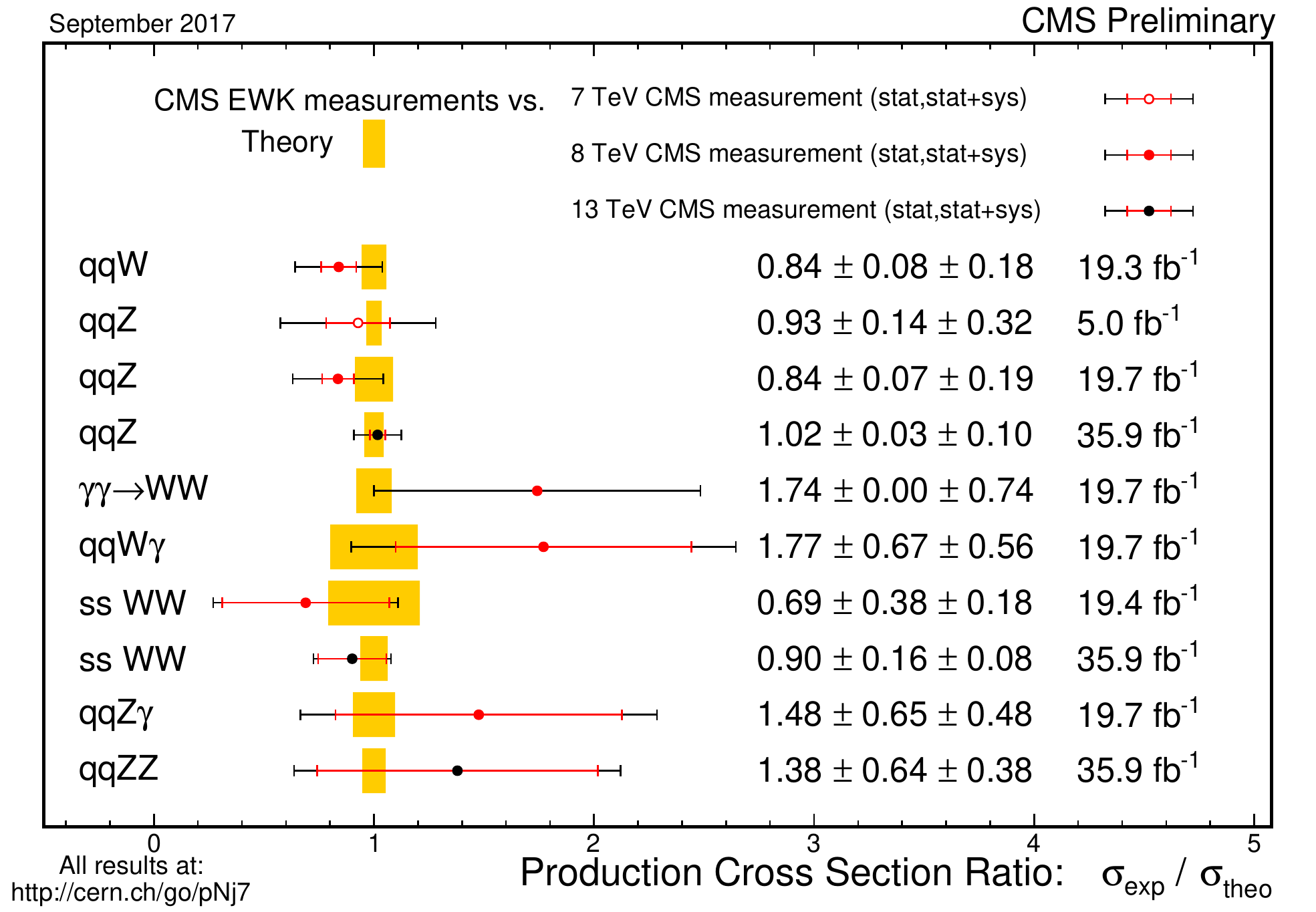}
\caption{(left) The dijet invariant mass spectrum in $Zjj$ events, where
  little hadronic activity is required in the rapidity region between
  the two tag-jets. The electroweak VBF process is clearly separated
  from the strong-interaction background at large
  $m_{jj}$. \cite{Aaboud:2017emo} (right) A summary for the electroweak
  production of single or dibosons by the CMS
  collaboration~\cite{summaryCMS}.}
\label{fig:lhc:vbfs}
\end{figure}

The electroweak production of one or two electroweak bosons in the VBF
and VBS processes, respectively, tests the self-interactions of three
or four electroweak bosons. Experimentally these are challenging
to measure due to small cross sections and large backgrounds from
processes mediated by the strong interaction, which are typically
separated from the signal through ``tagging'' the signature by a
rapidity gap and a high-mass dijet system~\cite{Bjorken:1992er} as
illustrated for the case of electroweak $Zjj$ production in
\Fref{fig:lhc:vbfs}(left). An overview of the status of the field is
shown in \Fref{fig:lhc:vbfs}(right). At this point in time, the
electroweak production of $W$ and $Z$ bosons has been observed with
high significance and studied in detail~\cite{Khachatryan:2014dea,
  Khachatryan:2016qkk, Aad:2014dta, Aaboud:2017fye,
  Aaboud:2017emo}. For VBS processes, most analyses have not yet
reached the observation stage. A significant exception is the VBS
production of same-sign $W^\pm W^\pm jj$, the golden VBS process due
to low strong-interaction backgrounds, where recently an observation
could be reported~\cite{Sirunyan:2017ret}.

\section{Status of the studies of the Higgs boson}
\label{sec:lhc:higgs}

\begin{figure}
  \includegraphics[width=0.46\linewidth]{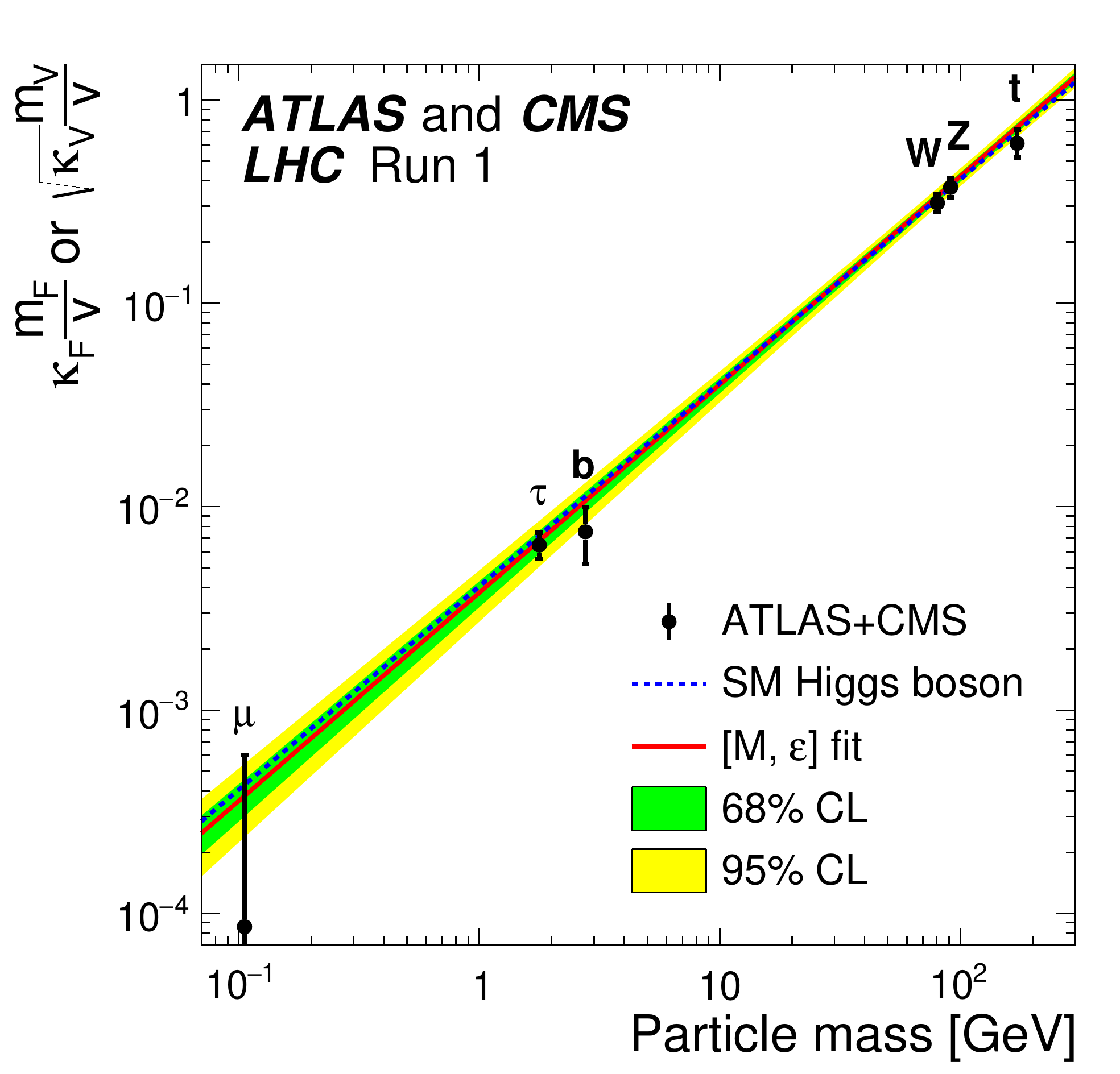}
  \includegraphics[width=0.54\linewidth]{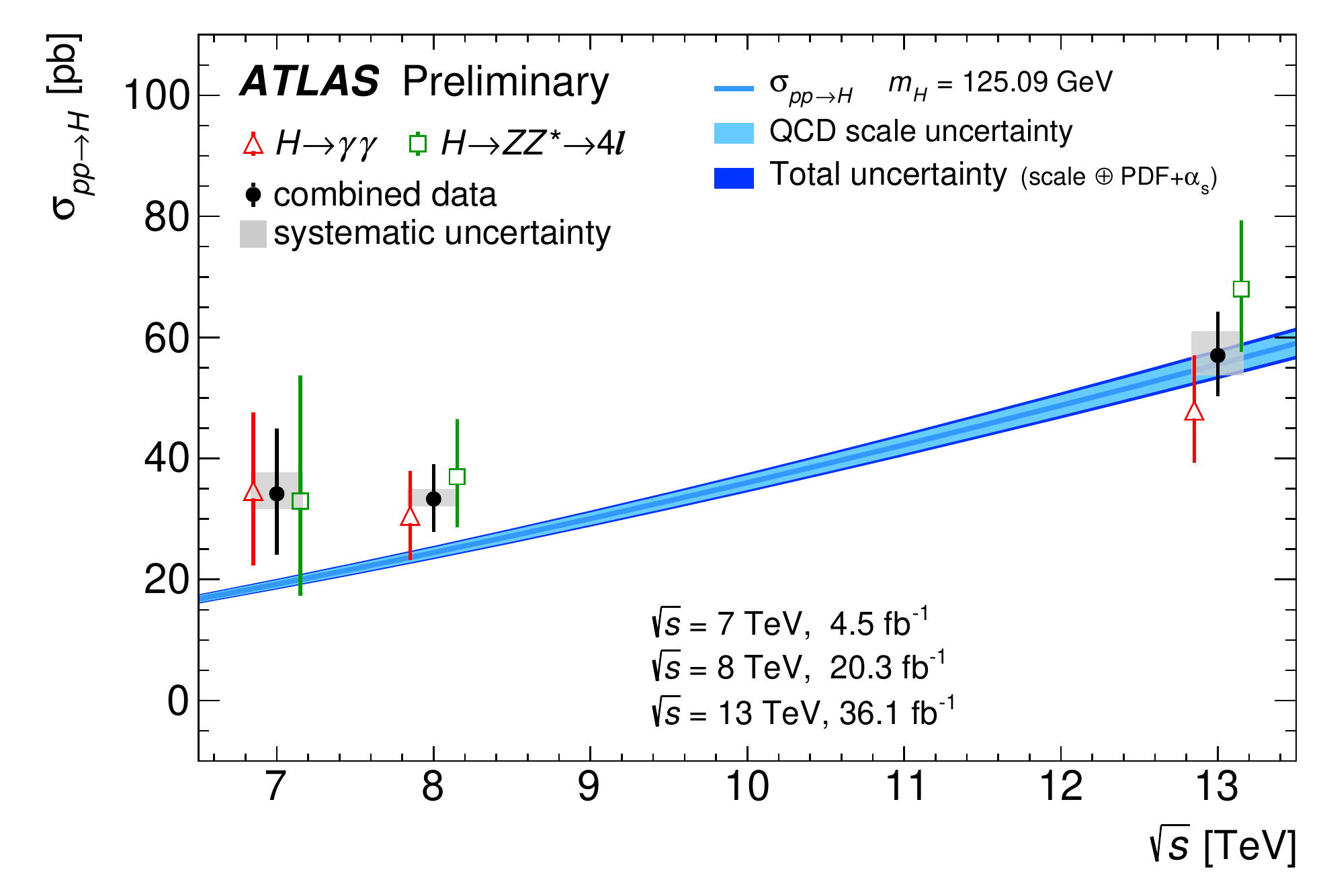}
  \caption{(left) Measurements of the coupling of the Higgs-boson to
    fermions and vector bosons as a function of the particle mass
    compared to the SM expectation.~\cite{Khachatryan:2016vau} (right)
    Combination of Higgs-boson production cross sections measured in
    the $H \to 4\ell$ and $H \to \gamma\gamma$ decay channels at
    centre-of-mass energies of $\sqrt{s} =7,8,13\TeV$~compared to the
    SM prediction~\cite{ATLAS-CONF-2017-047}.}
\label{fig:lhc:higgs}
\end{figure}

The Higgs boson was the last part of the Standard Model to be
discovered. Until then, the existence of the Higgs sector was seen as
``a mere conjecture'' by many, including Guido
Altarelli~\cite{Altarelli:2011vt}. Thanks to the well-working LHC and
its experiments, the discovery of ``a'' Higgs boson with a mass near
$125\GeV$ could be announced already in July 2012, delivering on the
promise that ``the LHC [was] designed to solve the Higgs
problem''~\cite{Altarelli:2011vt}. Since then, the properties have
been studied further using the data collected in Run 1 and now anew
using the Run 2 data. Already the Run 1 data allowed a precise
measurement of the mass to $m_H = 125.09 \pm
0.24\GeV$~\cite{Aad:2015zhl}, fixing the last free parameter of the
SM. Also the rates for the main production and decay rates could be
confirmed to be as predicted by the SM within a typical experimental
precision of $20$--$50\%$.  The overall agreement of Higgs production
and decay with the SM expectation may be summarised with the signal
strength of $\mu = 1.09 \pm 0.11$~\cite{Khachatryan:2016vau}, well
compatible with the SM expectation of $1$.  The quantum numbers were
confirmed to be $J^{CP} =
0^{++}$~\cite{Khachatryan:2014kca,Aad:2015mxa} with high
probability. As shown \Fref{fig:lhc:higgs}(left), the Run 1 data
established a reasonable agreement with the expectation, that the
masses of fermions and bosons are indeed given by Yukawa couplings
proportional to the particle masses, although also here the
experimental precision was typically just a few $10\%$.

As commented by Guido Altarelli, the good agreement of the
measurements with the minimal version of the SM and the
Englert-Brout-Higgs mechanism and at the same time an absence of other
evidence for physics beyond the Standard Model, poses new
theoretical challenges, as the Higgs boson is ``simple yet so
unnatural''~\cite{Altarelli:2013lla}. One of the paths to elucidate
the mechanism of electroweak symmetry breaking further are precision
measurements of the properties of the Higgs boson. With the
availability of the Run 2 data, Higgs measurements have entered now a
new phase. Exploiting the rarer but well-reconstructed decays $H \to
ZZ^* \to 4\ell$ and $H \to \gamma\gamma$ of just one experiment, the
total production cross section is already now measured at a precision
of about $10\%$ as shown in \Fref{fig:lhc:higgs}(right), which is similar
to the full Run 1 combination of all ATLAS and CMS data.
Very recently, ATLAS and CMS have been able to report 
evidence for the decay $H \to b\bar{b}$~\cite{Aaboud:2017xsd, Sirunyan:2017elk}
and for Higgs-boson production in association with top-quark pairs
$t\bar{t}H$~\cite{Aaboud:2017jvq,Sirunyan:2018shy}. At this
stage the results are compatible with the SM predictions with
a precision of $30-40\%$.
These analyses are very complex and make use of the full abilities of the detectors,
as well as mondern simulation and analysis techniques to control the large backgrounds.

In addition to more data, the precise measurements of Higgs-boson
properties depend critically on a further improved understanding of
the QCD aspects, both for the signal and background processes. A
recent highlight was the completion of the N$^3$LO calculation for the
dominant gluon-fusion process~\cite{Anastasiou:2015ema}. While this
significantly reduces the theoretical uncertainty in this particular
calculation, uncertainties from the gluon PDF, the strong coupling
\alphas, as well as the missing N$^3$LO corrections to the PDF
evolution and DIS, DY and other relevant cross sections will be a
challenge for the next years.

\section{Conclusion}

Since the year 2010 the LHC and the large experimental collaborations
have successfully delivered a detailed picture of the fundamental
constituents of matter and interactions at the electroweak scale and
beyond into the TeV range. As promised, the LHC solved the ``Higgs
problem'': the last missing piece of the Standard Model was discovered
and the work is ongoing to measure the properties of the newly
discovered particle. The Standard Model as we know it today is often
discussed as something that is ``unnatural'', just a ``low energy
effective theory'' and that ``just describes 5\% of the energy content
of the universe'', yet it is ``excessively'' successful. Guido
Altarelli was hopeful that there could be a fundamental break-through
out of this ``puzzling situation''. Meanwhile we should also
celebrate the achievements of the SM which was built in decades of systematic
work of theorists and experimentalists and which is able to describe the
diverse LHC data.

The work ahead for the LHC community is certainly a challenge, but
also full of exciting possibilities. The abundant and rich data
provided by the LHC collisions can and will be analyzed in diverse
ways to deepen our understanding of the strong and electroweak
interactions. In many measurements and searches we enter the
``precision regime'' as we accumulate large data sets and continue to
improve the understanding of our detectors. The methodology to
understand the proton structure and calculate predictions in QCD have
advanced to reach the percent level, which is truly remarkable
given the inherent complexity.

Guido Altarelli was often skeptical what concerned the theoretical
accuracy of QCD predictions which we experimentalists rely on so
critically and which we love to challenge. It would be interesting to
learn Guido's reaction to achievements such as the first measurement
of the $W$ boson mass at the LHC, which indeed requires a control of many
QCD effects at unprecedented level and skepticism is in order. It is
vital that we stay alert to the limitations of the theory.

\section*{Acknowledgements}

I would like to thank Stefano Forte, Aharon Levy and Giovanni Ridolfi
for taking the initiative for this book in memory and honour of Guido
Altarelli, for inviting me to contribute, as well as for giving valuable feedback
on the text.

\bibliographystyle{JHEP}
\bibliography{smlhc}

\end{document}